\documentclass[
reprint,
superscriptaddress,
nobibnotes,
amsmath,amssymb,
aps,
prmaterials,
citeautoscript
]{revtex4-2}

\usepackage{graphicx}
\usepackage{dcolumn}
\usepackage{bm}
\usepackage{hyperref}
\hypersetup{
    breaklinks=true,
    bookmarks=true,
    pdfauthor={},
    pdftitle={},
    colorlinks=true,
    citecolor=blue,
    urlcolor=blue,
    linkcolor=magenta,
    pdfborder={0 0 0}
}

\usepackage{mathtools} 
\usepackage{upgreek} 

\usepackage{color,soul} 
\usepackage[normalem]{ulem} 
\usepackage{xfrac} 
\usepackage{braket} 
\usepackage{cases} 
\usepackage{placeins} 
\usepackage{booktabs} 

\usepackage{floatrow} 
\usepackage[caption=false]{subfig}  

\usepackage{xr} 
\begin{document}

\floatsetup[figure]{style=plain,subcapbesideposition=top} 
\floatsetup[table]{style=plain, capposition=top} 

\preprint{APS/123-QED}

\title{A CN complex as an alternative to the T center in Si}


\author{J. K. Nangoi}
\email[Corresponding author: nangoi@ucsb.edu]{}
\affiliation{
    Materials Department, University of California, Santa Barbara, California 93106, USA
}

\author{M. E. Turiansky}
\affiliation{
    Materials Department, University of California, Santa Barbara, California 93106, USA
}
\affiliation{
    US Naval Research Laboratory, Washington, DC 20375, USA
}

\author{C. G. Van de Walle}
\affiliation{
    Materials Department, University of California, Santa Barbara, California 93106, USA
}


%


\begin{abstract}

    We present a first-principles study of a carbon-nitrogen (CN) impurity complex in silicon as an isoelectronic alternative to the T center 
    [(CCH)$_\mathrm{Si}$]. The latter has been pursued for applications in quantum information science, yet 
    its sensitivity to the presence of hydrogen is still problematic.
    Our proposed complex has no hydrogen, thereby eliminating this issue. 
    First, we show that the CN complex is stable against decomposition into substitutional and interstitial defects. 
    Next, we show that due to being isoelectronic to the T center, the CN complex has a similar electronic structure, and therefore could be used in similar applications. 
    We assess several low-energy configurations of the CN complex, finding (CN)$_\mathrm{Si}$ to be stable and have the largest Debye-Waller factor. 
    We predict a zero-phonon line (ZPL) of 828~meV (in the telecom S-band) and a radiative lifetime of 4.2~$\mu$s, 
    comparable to the T center. 
    Due to the presence of a bound exciton, choice of the exchange-correlation functional and also supercell-size scaling of the ZPL and transition dipole moment require special scrutiny; we 
    rigorously justify our extrapolation schemes that allow computing values in the dilute limit.

\end{abstract}


\maketitle




Point defects in semiconductors or insulators are being studied for quantum information science applications, including as spin qubits for quantum computing~{\cite{ref:bradley2019, ref:yan2021, ref:simmons2024}} and single-photon emitters for quantum networks~{\cite{ref:ruf2021, ref:yan2021, ref:simmons2024}}.
For the former, the spin coherence time is the main metric, as it determines how long quantum information remains coherent.
For networking, single photons carry the quantum information~{\cite{ref:northup2014, ref:turiansky_rational}}.
A high Debye-Waller (DW) factor is desired so that most photons are emitted into the zero-phonon line (ZPL), for which the photons are in a well-defined quantum state~{\cite{ref:turiansky_rational}}.

\begin{figure}[h!]
    \centering
    \includegraphics[width=\linewidth]{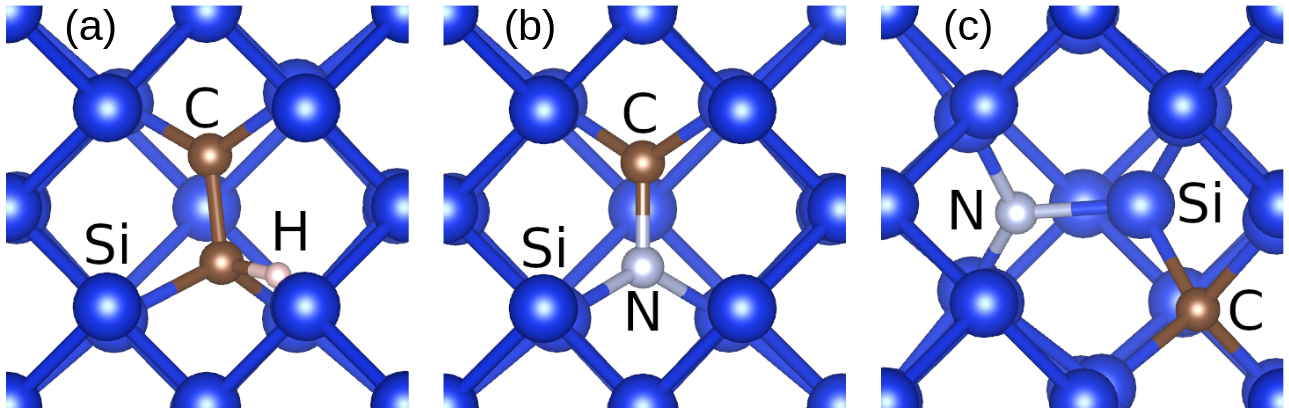}

    \caption{
        Structure of (a) T center, (CCH)$_\mathrm{Si}$, (b) (CN)$_\mathrm{Si}$, and (c) C$_\mathrm{Si}$(NSi)$_\mathrm{Si}$.
    }

    \label{fig:struct}
\end{figure}

The nitrogen-vacancy (NV) center in diamond has been studied extensively for these applications.
It has a triplet ground state with spin coherence time exceeding milliseconds~{\cite{ref:weber2010}},
but a DW factor of only $\sim$3\%~{\cite{ref:davies81}}.
Alternatives to the NV center have been studied, such as the silicon-vacancy center in diamond~\cite{ref:bradac2019}, carbon-vacancy and silicon-vacancy in cubic boron nitride~{\cite{ref:turiansky-cubic-bn}}, or color centers in silicon~\cite{ref:bergeron, ref:dhaliah, ref:simmons2024}.
Silicon is an attractive host material because it offers the prospect of integration with Si-based electronics~{\cite{ref:bergeron}}; it is also far easier to grow and process than diamond~{\cite{ref:weber2010}}.
The T center in Si, a complex in which two carbons and one hydrogen substitute on a Si site [denoted by (CCH)$_\mathrm{Si}$, Fig.~{\ref{fig:struct}}(a)], has been intensively pursued experimentally due to its high spin coherence times, 
large DW factor, 
and emission in the O-band of telecom wavelengths~\cite{ref:bergeron, ref:macquarrie, ref:higginbottom22, ref:higginbottom23, ref:deabreu, ref:islam}.

Despite the promise, formation and stability of the T center remain a concern; it was found to be 
``prone to (de)hydrogenation and so requires very precise annealing conditions (temperature and atmosphere) to be efficiently formed''~\cite{ref:dhaliah}. 
Identifying alternatives to the T center, preferably without H, would thus be beneficial.

In this letter, we propose a carbon-nitrogen complex as an analog of the T center. 
Replacing the hydrogen and one carbon atom in (CCH)$_\mathrm{Si}$ with a nitrogen atom keeps the center isoelectronic with the T center, since N contains the same number of electrons and protons as C+H.
The absence of H makes this complex more stable with regard to (de)hydrogenation. 
Like H, N has non-zero nuclear spin and can thus be exploited to store quantum information.
We demonstrate the stability of this center by calculating the formation and decomposition energies, and thoroughly assess its electronic structure and optical properties, including 
the Huang-Rhys/DW factors, 
energy of the ZPL transition, 
and radiative lifetime. 
Our results show that the CN center is a 
promising candidate 
for applications in quantum information science.


Our first-principles studies are based on density-functional theory (DFT)
with projector-augmented wave (PAW) potentials~\cite{ref:paw, ref:vasp-paw} as implemented in the Vienna Ab initio Simulation Package (VASP)~\cite{ref:vasp, ref:vasp2}, with a plane-wave cutoff of 400~eV.
We use the hybrid functional of Heyd, Scuseria, and Ernzerhof (HSE)~\cite{ref:HSE03, ref:HSE06} with the default mixing parameter of 25\%; for select results (as described below) we also employ the PBE0 (Perdew, Burke, and Ernzerhof) hybrid functional~\cite{ref:perdew1996_pbe0, ref:ernzerhof1999_pbe0, ref:adamo1999_pbe0} with a mixing parameter of 13.6\%.
For the primitive cell with a Brillouin-zone sampling mesh of $11 \times 11 \times 11$, we find an HSE lattice constant of 5.433~{\AA} and a band gap of 1.15~eV, both in agreement with experiment (5.431~{\AA}~\cite{ref:si_exp-lattConst} and 1.17~eV at 0 K~{\cite{ref:si_exp_gap})}.
We model the defect in a supercell geometry, using the $\Gamma$ point to sample the Brillouin zone. 
Most of our results are presented for a 512-atom supercell ($4 \times 4 \times 4$ multiple of the conventional cubic cell), 
allowing us to make direct comparison with previous work~{\cite{ref:dhaliah}}. We also use up to 1000-atom supercells ($5 \times 5 \times 5$) to better describe the bound excitons that are present in the excited states of the centers studied here.
Structural optimizations are performed until the forces are less than 0.01~eV/\AA.


Previous first-principles calculations identified two possible configurations of the CN complex: (1) a C-N split interstitial (CN)$_\mathrm{Si}$~{\cite{ref:platonenko_N-int}} [Fig.~{\ref{fig:struct}}(b)], which can be thought of as replacing C-H with N in the T center (CCH)$_\mathrm{Si}$, and (2) a complex of a substitutional C atom and a N-Si split interstitial, C$_\mathrm{Si}$(NSi)$_\mathrm{Si}$~{\cite{ref:kuganathan2023, ref:sgourou2024}} [Fig.~{\ref{fig:struct}}(c)]. 

\begin{figure}[h!]
    \centering
    \includegraphics[width=0.95\linewidth]{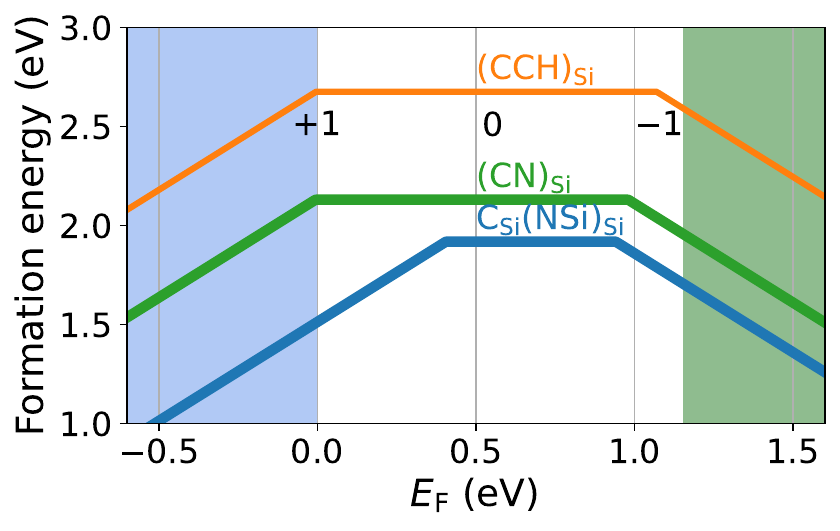}

    \caption{
        Defect formation energies as a function of Fermi level for the T center (orange), 
        (CN)$_\mathrm{Si}$ (green), 
        and C$_\mathrm{Si}$(NSi)$_\mathrm{Si}$ (blue). 
        Blue and green shades indicate valence and conduction bands. 
        The valence-band maximum (VBM) is set to 0, and the conduction-band minimum (CBM) is at 1.15~eV, our calculated Si band gap.
    }

    \label{fig:Eform}
\end{figure}

Figure~\ref{fig:Eform} shows our calculated formation energies $E^f$ as a function of the Fermi level for both CN structures, and also for the T center. $E^f$ is given by~{\cite{ref:vdw-review}}:
\begin{align}
    E^f [X^q] = E\textsubscript{tot} [X^q] - E\textsubscript{tot} [\mathrm{Si}] - \sum_i n_i \mu_i + q E_F + E_\textsubscript{corr}, 
    \nonumber
\end{align}
where $E\textsubscript{tot} [X^q]$ is the total energy of the supercell containing the defect $X$ in charge state $q$, $E\textsubscript{tot} [\mathrm{Si}]$ is the total energy of the equivalent supercell containing perfect host material, $n_i$ is the number of atoms of type $i$ added to ($n_i >0$) or removed from ($n_i<0$) the supercell, $\mu_i$ is the chemical potential of atom type $i$, $E_F$ is the Fermi level, and $E_\textsubscript{corr}$ 
is a finite-size correction for charged defects~\cite{ref:fnv}. 
For the $\mu_i$ we use the total energies per atom of bulk Si, diamond, H$_2$, and N$_2$.

Our results for the T center agree with Ref.~\onlinecite{ref:dhaliah}. 
The +1 charge state of the T center and (CN)$_\mathrm{Si}$ are not stable in the gap, as seen in Fig.~{\ref{fig:Eform}}. 
We find that the $-$1 charge state of the T center, (CN)$_\mathrm{Si}$, and C$_\mathrm{Si}$(NSi)$_\mathrm{Si}$, and also the +1 charge state of C$_\mathrm{Si}$(NSi)$_\mathrm{Si}$, all have net zero spin 
(singlet) (see Fig.~{\ref{fig:KS-eigenvals_pm1}} in the Supplemental Material (SM)~{\cite{ref:supplemental}}), precluding their use as spin qubits. 
We also note that the excited state of, e.g., the $-$1 charge state involves exciting an electron to the conduction-band minimum (CBM), leaving a neutral defect center behind. The electron will thus feel no Coulomb attraction 
and hence cannot act as a single-photon emitter.  
Similar arguments apply to the +1 charge state. 
These charge states are therefore less useful for quantum information applications.
We thus focus on the neutral charge state, which is stable over the majority of the band gap for (CN)$_{\rm Si}$, with the negative charge state occurring only for Fermi levels within 0.17~eV of the CBM. [C$_\mathrm{Si}$(NSi)$_\mathrm{Si}]^0$ is stable for Fermi levels between 0.41 and 0.94~eV.

The stability of alternate atomic configurations in the case of the CN center inspired us to investigate whether the CCH center could also be stable in other structures; we found these to be 0.6--2.5~eV higher in energy than the accepted (CCH)$_\mathrm{Si}$ structure of the T center (see Sec.~\ref{sec:supp_other-defects} of the SM~\cite{ref:supplemental}). 

We assess the stability of the CN defects [both C$_\mathrm{Si}$(NSi)$_\mathrm{Si}$ and (CN)$_\mathrm{Si}$] with respect to decomposition into constituent defects 
(substitutions and interstitials) 
by calculating the decomposition energy $\Delta E^f \equiv \sum E^f [\mathrm{products}] - E^f~[\mathrm{CN~defect}]$; 
a positive energy indicates the reaction is endothermic. 
For C and N interstitials, the split-interstitial configurations are found to be lowest in energy, consistent with Refs. {\onlinecite{ref:tersoff1990, ref:platonenko_N-int, ref:platonenko_C-sub, ref:platonenko_C-int, ref:simha2023}
(see Sec.~{\ref{sec:supp_other-defects}} in the SM~{\cite{ref:supplemental}} for atomic structures and formation energies). 

Table~{\ref{tab:decomp-energies}} shows our results for decomposition energies, taking into account that the overall charge state should remain neutral.
All decomposition energies are positive, meaning both configurations are stable against all considered decompositions. 
The $\Delta E^f$ for (CN)$_\mathrm{Si}^0$ are $\sim$0.2~eV lower than those for [C$_\mathrm{Si}$(NSi)$_\mathrm{Si}$]$^0$, because the formation energy of (CN)$_\mathrm{Si}^0$ is slightly higher than that of [C$_\mathrm{Si}$(NSi)$_\mathrm{Si}$]$^0$ (Fig.~{\ref{fig:Eform}}). 
Decomposition into C$_\mathrm{Si}^0$ and (NSi)$_\mathrm{Si}^0$ has the smallest energy, with
a value that is comparable to the lowest decomposition energy (0.80~eV) 
for (CCH)$_\mathrm{Si}$ calculated in Ref.~{\onlinecite{ref:dhaliah}}. 
Furthermore, as mentioned above, the absence of hydrogen in the CN defect is advantageous.

We use the climbing image nudged elastic band method~{\cite{ref:Henkelman2000_climbing-neb}} to calculate the migration barrier of (NSi)$_\mathrm{Si}^0$ (the most mobile constituent), resulting in 0.68~eV (see Sec.~{\ref{sec:supp_neb}} in the SM~{\cite{ref:supplemental}}). 
We can thus estimate the barrier height of the lowest-energy decomposition reaction to be 1.50~eV for [C$_\mathrm{Si}$(NSi)$_\mathrm{Si}$]$^0$ and 1.29~eV for (CN)$_\mathrm{Si}^0$. 

\begin{table}[h!]
\caption{
    Decomposition energies $\Delta E^f$ for 
    (CN)$_\mathrm{Si}^0$ and [C$_\mathrm{Si}$(NSi)$_\mathrm{Si}$]$^0$. 
    %
    \label{tab:decomp-energies}
}
\begin{ruledtabular}
\begin{tabular}{lcc}
    & \multicolumn{2}{c}{$\Delta E^f$ (eV)} \\
    \hline
    & (CN)$_\mathrm{Si}^0$ & [C$_\mathrm{Si}$(NSi)$_\mathrm{Si}$]$^0$ \\
    \hline
    $\to$ C$^{ 0}_\mathrm{Si}$ + (NSi)$^{ 0}_\mathrm{Si}$ & 0.61 & 0.82 \\
    $\to$ C$^{-1}_\mathrm{Si}$ + (NSi)$^{+1}_\mathrm{Si}$ & 1.33 & 1.54 \\
    $\to$ C$^{+1}_\mathrm{Si}$ + (NSi)$^{-1}_\mathrm{Si}$ & 1.69 & 1.90 \\
    $\to$ (CSi)$^{ 0}_\mathrm{Si}$ + N$^{ 0}_\mathrm{Si}$ & 3.12 & 3.33 \\
    $\to$ (CSi)$^{-1}_\mathrm{Si}$ + N$^{+1}_\mathrm{Si}$ & 3.65 & 3.86 \\
    $\to$ (CSi)$^{+1}_\mathrm{Si}$ + N$^{-1}_\mathrm{Si}$ & 3.70 & 3.91 \\
\end{tabular}
\end{ruledtabular}
\end{table}


We now study the electronic structure by analyzing the spin-polarized Kohn-Sham states and wavefunctions in both the ground and excited electronic states. 
Figure~\ref{fig:KS-eigenvals} compares (CN)$_\mathrm{Si}^0$ and [C$_\mathrm{Si}$(NSi)$_\mathrm{Si}$]$^0$ with (CCH)$_\mathrm{Si}^0$
(the T center, for which our results agree with Ref.~\onlinecite{ref:dhaliah}).
In the ground state of the T center, the $a''$ antibonding state associated with the $C_{1h}$ symmetry undergoes exchange splitting: the occupied spin-up state is below the VBM and the unoccupied spin-down state lies just below the CBM. 
The ground states of the CN centers are qualitatively similar, with exchange splitting in the $b$ antibonding state for (CN)$_\mathrm{Si}$ ($C_{2v}$ symmetry) and in the $a$ state for C$_\mathrm{Si}$(NSi)$_\mathrm{Si}$ ($C_1$ symmetry), although the occupied spin-up $a$ state is now above the VBM. 
The unoccupied states for all 3 defects [Figs.~{\ref{fig:chgDens_sup4}}(a)--(c)] and the occupied state for C$_\mathrm{Si}$(NSi)$_\mathrm{Si}$ [Figs.~{\ref{fig:chgDens_sup4}}(d)] are localized at the defect site.

\begin{figure}[h!]
    \centering

    \includegraphics[width=\linewidth]{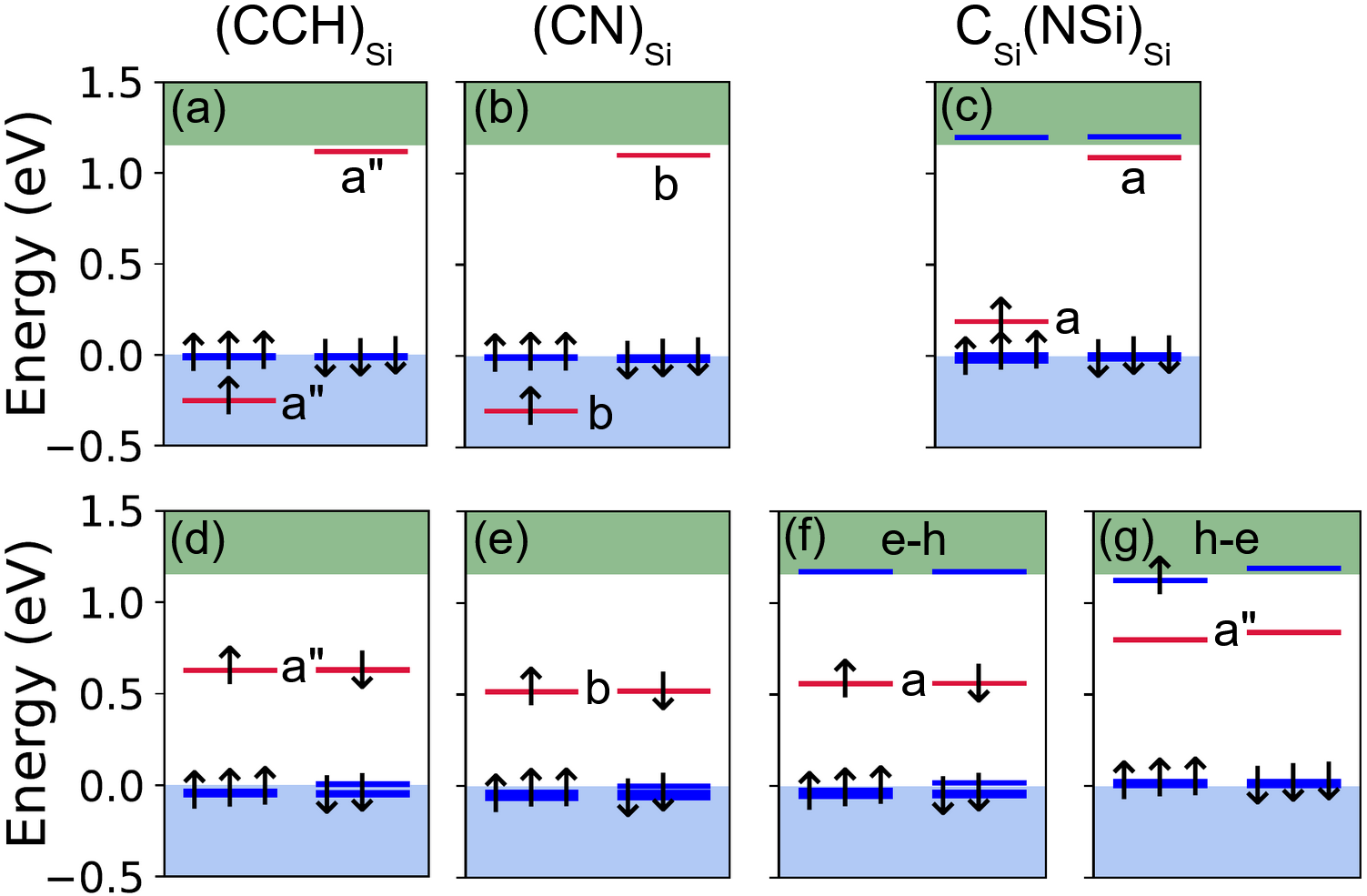}
    \caption{
        Ground- (top row) and excited-state (bottom row) Kohn-Sham states for the neutral charge state. 
        ``e-h'' (``h-e'') labels the localized-electron (hole) case. 
        Blue and green shades indicate valence and conduction bands. 
        Red levels are the states associated with the defect, and blue levels are valence- and conduction-band states.
    }
    
    \label{fig:KS-eigenvals}
\end{figure}

\begin{figure}[h!]
    \centering
    
    \includegraphics[width=\linewidth]{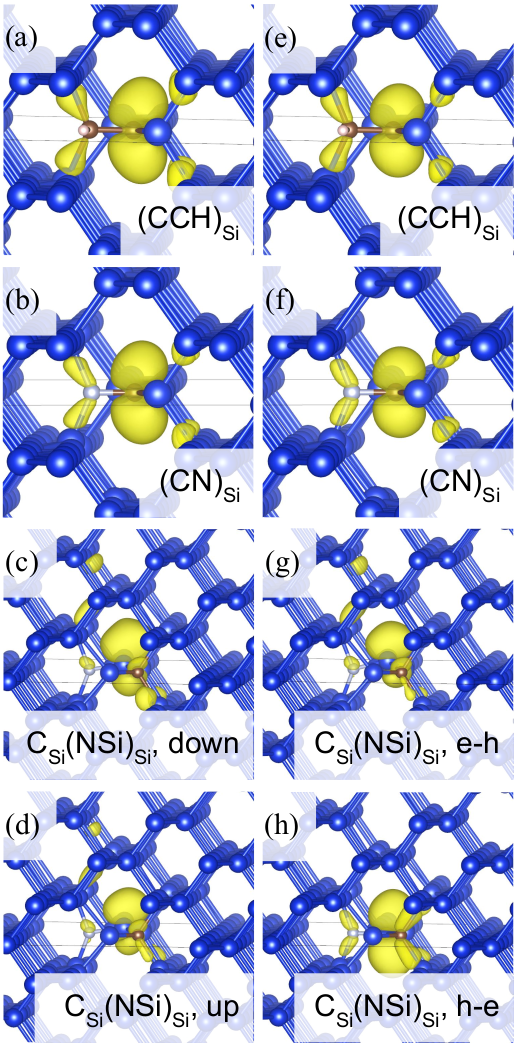}    
    \caption{
        Isosurfaces (yellow) of real-space Kohn-Sham probability densities for the neutral charge state 
        of the ground-state in-gap empty spin-down state [(a)--(c)] and filled spin-up state [(d)], 
        and 
        of the excited-state in-gap filled spin-down state [(e)--(g) for localized-electron case] and empty spin-up state [(h) for localized-hole case]. 
        Blue circles are Si; brown, C; light blue, N; and pink, H.
    }

    \label{fig:chgDens_sup4}
\end{figure}

To model excited states, we use the  constrained occupation $\Delta$SCF approach~{\cite{ref:constr-occ-delta-scf}}, where we excite a spin-down electron from the VBM to the unoccupied defect state, constrain the electron occupation, and reoptimize the structure. 
The resulting structure has an in-gap state with both spin channels occupied [Figs.~{\ref{fig:KS-eigenvals}(d)--(f)] that is localized at the defect site [Figs.~{\ref{fig:chgDens_sup4}(e)--(g)}] (maintaining the symmetry of the ground state), and leaves a hole with hydrogenic nature bound to the localized electron.
The localized electron and the hole form a bound exciton, where, 
within hydrogenic effective mass theory, 
the hole 
approximately 
has an effective Bohr radius of $\sim$13~{\AA} 
as shown in Sec.~{\ref{sec:supp_sup-dep}} of the SM~\cite{ref:supplemental}. 

For C$_\mathrm{Si}$(NSi)$_\mathrm{Si}$, 
it is also possible to excite a spin-up electron 
from the in-gap defect state into the CBM, resulting in an exciton with 
a localized \emph{hole} and a hydrogenic \emph{electron} [Figs.~{\ref{fig:KS-eigenvals}}(g) and \ref{fig:chgDens_sup4}(h)]. 
The corresponding effective Bohr radius for the electron is approximately $\sim$6~{\AA} (see Sec.~{\ref{sec:supp_sup-dep}} of the SM~\cite{ref:supplemental}).
In contrast, for both the T center and (CN)$_\mathrm{Si}$, exciting the electron from the 
$a''$ or $b$ state below the VBM into the CBM results in an excited state where 
the hole is at the VBM 
instead of being localized at an in-gap defect state.

We note that the localized-hole excited state of C$_\mathrm{Si}$(NSi)$_\mathrm{Si}$ has $C_{1h}$ symmetry, unlike the $C_1$ symmetry for both the ground state and the localized-electron excited state, as apparent in the charge density [Fig.~{\ref{fig:chgDens_sup4}}(h)]. 
This is because, as seen in the figure, the C, N, and Si bonded to both C and N now form a mirror plane, unlike in the ground state [Fig.~{\ref{fig:chgDens_sup4}}(c) and (d)] and the localized-electron excited state [Fig.~{\ref{fig:chgDens_sup4}}(g)].

Table~{\ref{tab:huang-rhys}} shows our calculated Huang-Rhys factors $S = E_\mathrm{r} / (\hbar \Omega)$ and DW factors $\exp{(-S)}$; $E_\mathrm{r}$ and $\Omega$ are the relaxation energy and phonon frequency in the electronic ground state within the one-dimensional approximation~{\cite{ref:huang-rhys-orig, ref:alkauskas2012_huang-rhys, ref:alkauskas2014_hr-dw}}.
Details are in Sec.~{\ref{sec:supp_HR}} of the SM~{\cite{ref:supplemental}}. 
Our calculated DW factor for the T center is $\sim$9\%; Ref.~{\onlinecite{ref:xiong_hautier}} reports a calculated value of 16.5\%, while the value obtained from photoluminescence (PL) data is 23\%~{\cite{ref:bergeron}}. 
The discrepancy with Ref.~{\onlinecite{ref:xiong_hautier}} could be due to differences in $E_{\mathrm{r}}$ or $\Omega$.
The discrepancy with experiment could be because our DW factors are based on a one-dimensional model, which underestimates the more realistic, multi-dimensional model~{\cite{ref:turiansky2025}}, 
so we focus here on {\it trends}. 
The DW factors for C$_\mathrm{Si}$(NSi)$_\mathrm{Si}$ are below 1\%, limiting their usefulness for single-photon emitters.
In the following we focus on (CN)$_\mathrm{Si}$. 

\begin{table}[h!]
\caption{
    Relaxation energies ($E_r$), ground-state phonon frequencies ($\Omega$), and Huang-Rhys ($S$) and Debye-Waller (DW) factors. 
    %
    \label{tab:huang-rhys}
}
\begin{ruledtabular}
\begin{tabular}{lcccc}
    & $E_r$ (eV) & $\hbar\Omega$ (meV) & S & DW (\%) \\ 
    \hline
    (CCH)$_\mathrm{Si}$                    & 0.079 & 33 & 2.4   &  9.2\\
    (CN)$_\mathrm{Si}$                     & 0.107 & 35 & 3.0   &  5.0 \\
    C$_\mathrm{Si}$(NSi)$_\mathrm{Si}$ e-h & 0.157 & 28 & 5.6   &  0.4 \\
    C$_\mathrm{Si}$(NSi)$_\mathrm{Si}$ h-e & 0.297 & 28 & 10.5 &  0.003 \\ 
\end{tabular}
\end{ruledtabular}
\end{table}

The energy of the ZPL transition $E_\mathrm{ZPL}$ is calculated as the total-energy difference between the excited state and the ground state.
Using a 512-atom supercell, we find $E_\mathrm{ZPL}$ = 981~meV for the T center, in agreement with Ref.~\onlinecite{ref:dhaliah}. 
However, as seen in Fig.~{\ref{fig:supdep_ZPL}} the calculated $E_\mathrm{ZPL}$ depends on the supercell size because our supercells are not large enough to completely fit the hydrogenic wavefunctions (Fig.~{\ref{fig:chgDens_bound}} in the SM {\cite{ref:supplemental}})~\footnote{Supercell-size dependence for the T center was also observed in Ref.~{\onlinecite{ref:alaerts2025}}, which calculated exciton binding energies 
and dipole moment changes}. 
The dependence of $E_\mathrm{ZPL}$ on supercell size is well described by a linear fit to the inverse of the supercell volume (Fig.~{\ref{fig:supdep_ZPL}}).
An extrapolation to the dilute limit based on our HSE values produces a value of 1064~meV, overestimating the experimentally measured ZPL of 935~meV~\cite{ref:bergeron}. 

\begin{figure}[h!]
    \centering
    \includegraphics[width=\linewidth]{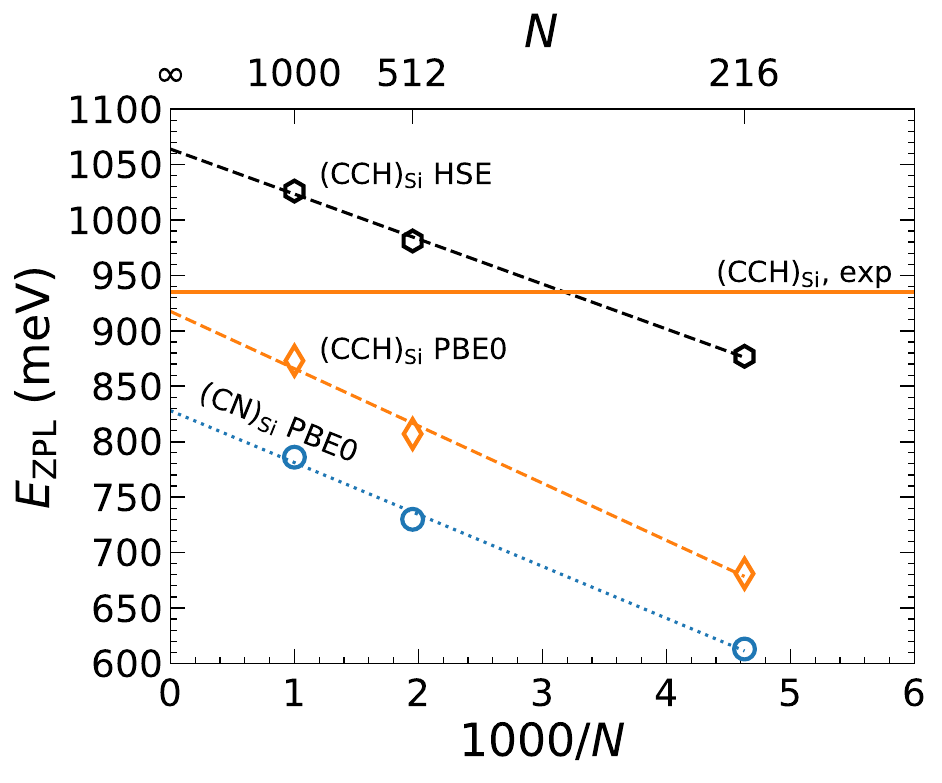}
 \caption{
       Zero-phonon line energies for the T and CN centers, calculated using either single-shot PBE0 or HSE and plotted as a function of inverse supercell size ($N$ is the number of atoms in the defect-free supercell).
       The lines are linear fits. 
       The horizontal orange line represents the measured ZPL of the T center, 935~meV from Ref.~{\onlinecite{ref:bergeron}}.
    }
    \label{fig:supdep_ZPL}
\end{figure}

An investigation of the cause of this deviation indicated that the ZPL values are sensitive to the DFT functional.
We found that PBE0 gives a better extrapolated ZPL for the T center, and  therefore choose it to improve predictions for the CN defects. 

A PBE0 mixing parameter of 13.6\% yields a lattice constant of 5.446~{\AA} and a band gap of 1.23~eV, which reproduces the experimental $T$=0 gap of 1.17~eV {\cite{ref:si_exp_gap}} plus the zero-point renormalization of 60~meV~{\cite{ref:karaiskaj2002_si-zpr}}. 
We found that PBE0 geometries are very close to those obtained with HSE, and hence we use single-shot PBE0 (i.e., only performing electronic structure optimization) using the relaxed HSE geometries (scaled according to the slight difference in lattice constant). 
The accuracy of this approach is demonstrated in 
Table~{\ref{tab:supp_PBE0-full-vs-single-shot}} 
of the SM~\cite{ref:supplemental}.

As shown in Fig.~{\ref{fig:supdep_ZPL}}, extrapolation of the PBE0 values to the dilute limit yields 918~meV for the T center [(CCH)$_\mathrm{Si}$],
significantly closer to experiment (935~meV, Ref.~{\onlinecite{ref:bergeron}}) than the HSE value of 1064 meV.

For (CN)$_\mathrm{Si}$, we find an extrapolated PBE0 value of 828~meV (in the S-band of the telecom wavelength region~{\cite{ref:telecom-region}}); 
see Sec.~{\ref{sec:supp_ZPL}} for the values for C$_\mathrm{Si}$(NSi)$_\mathrm{Si}$.
A number of luminescence lines with comparable energies have been observed in Si, at 829.8~meV~{\cite{ref:davies89}}, 844~meV~{\cite{ref:conzelmann1983}}, and 856~meV; the latter was suggested to be associated with interstitial C{~\cite{ref:davies89}}. 
Experimental ZPLs ranging from 746 to 772~meV have been attributed to complexes containing C and N~\cite{ref:dornen85, ref:dornen86-prb, ref:dornen86, ref:dornen87, ref:dornen88, ref:davies89}; 
however, no reliable microscopic identification of the structure of these complexes is available, 
and the cited references indicate that some may involve additional impurities such as oxygen. 

Finally, we calculate the radiative lifetime $\tau$ of the ZPL transition using the Weisskopf-Wigner formula~\cite{ref:stoneham, ref:wig-weiss, ref:turiansky_rational},
\begin{align}
    \frac{1}{\tau} &= \left( \frac{\mathcal{E}_\mathrm{eff}}{\mathcal{E}_0} \right)^2 \frac{n_r}{3 \pi \epsilon_0 c^3 \hbar^4}  (E_\mathrm{ZPL})^3 |\mathbf{\mu}|^2, 
   \label{eqn:radiative-lifetime}
\end{align}
where $n_r$ is the refractive index of the host material (3.38 for Si~{\cite{ref:si_exp-lattConst}}), 
$c$ is the speed of light, 
and $\mu$ is the transition dipole moment (TDM). 
The prefactor $(\mathcal{E}_\mathrm{eff}/\mathcal{E}_0)^2$ accounts for local-field effects~\cite{ref:stoneham, ref:turiansky_rational}, which tend to increase the rate.
Here we take $\mathcal{E}_\mathrm{eff} \approx \mathcal{E}_0$, which is a common approximation~{\cite{ref:razinkovas, ref:turiansky_rational}}.

Calculating $\mu$ is challenging because it requires accurately describing the hydrogenic wavefunction corresponding to the bound exciton.
We therefore do not attempt to calculate $\mu$ in the excited state, but we approximate $\mu$ by scaling the TDM calculated in the \emph{ground} state, $\mu^0$, to account for the hydrogenic nature:
\begin{align}
    |\mathbf{\mu}|^2 \approx t |\mathbf{\mu}^0|^2. \label{eqn:main_mu-sq_approx}
\end{align}
Here $t = \tilde{V} / [\pi (a_0^*)^3]$, where $\tilde{V}$ is the supercell volume and $a_0^*$ is the effective Bohr radius of the hydrogenic wavefunction. For details, see Sec.~{\ref{sec:supp_mu}} of the SM~{\cite{ref:supplemental}}~\footnote{ The scaling method here results in a radiative lifetime formula that is is analogous to the radiative capture rate formalism~{\cite{ref:dreyer}} as explained in Sec.~{\ref{sec:supp_capture}} in the SM~{\cite{ref:supplemental}} }. 

Using the approximate $\mu$ 
and the extrapolated ZPLs, 
we find very similar radiative lifetimes of 
4.70~$\mu$s (which agrees well with 4.9~$\mu$s deduced from experiments in Ref.~{\onlinecite{ref:kazemi2025}}) and 4.18~$\mu$s for the T center and (CN)$_\mathrm{Si}$. 
(See Table~\ref{tab:supp_muSq-tau} for the values for C$_\mathrm{Si}$(NSi)$_\mathrm{Si}$.)


In conclusion, we propose the CN complex 
as a hydrogen-free alternative to the T center for similar quantum applications. 
We have shown that both (CN)$_\mathrm{Si}$ and  C$_\mathrm{Si}$(NSi)$_\mathrm{Si}$ are stable against decomposition to C and N substitutional/interstitial defects, and have electronic structures similar to the T center: a neutral charge state that is stable in the band gap, similar Kohn-Sham eigenvalues and eigenstates in both the ground and excited state, and an excited state consisting of a bound exciton with a localized electron and a hydrogenic hole. 
Additionally, C$_\mathrm{Si}$(NSi)$_\mathrm{Si}$ has an excited state consisting of a bound exciton with a localized \emph{hole} and a hydrogenic \emph{electron}. 
We carefully handle the supercell-size scaling for the ZPL and propose an extrapolation procedure for the radiative lifetimes, allowing us to obtain results in the dilute limit.
We find the lifetime for (CN)$_\mathrm{Si}$ to be similar to the (CCH)$_\mathrm{Si}$ center. 
These results, combined with the fact that the predicted ZPL of (CN)$_\mathrm{Si}$ is in the telecom S-band, render the (CN)$_\mathrm{Si}$ center a promising hydrogen-free alternative to the T center.

\FloatBarrier 


\begin{acknowledgements}

We gratefully acknowledge discussions with 
D. Waldh\"{o}r, W. Lee, M. W. Swift, C. A. Broderick, and Y. Chen. 
This work was supported by the U.S. Department of Energy (DOE), Office of Science (SC), National Quantum Information Science Research Centers, Co-design Center for Quantum Advantage (C$^2$QA) under contract number DE-SC0012704, and
used computing resources provided by 
the National Energy Research Scientific Computing Center (NERSC), a User Facility supported by the DOE SC under Contract No. DE-AC02-05CH11231 using NERSC award BES-ERCAP0021021 and BES-ERCAP0028497. 
Additional resources were provided by
the Texas Advanced Computing Center (TACC) at The University of Texas at Austin
and the San Diego Supercomputer Center (SDSC) Expanse at the University of California, San Diego 
through allocation DMR070069 from the Advanced Cyberinfrastructure Coordination Ecosystem: Services \& Support (ACCESS) program~\cite{ref:access}, which is supported by National Science Foundation grants \#2138259, \#2138286, \#2138307, \#2137603, and \#2138296. 

\end{acknowledgements}


\nocite{ref:tersoff1990, ref:platonenko_N-int, ref:platonenko_C-sub, ref:platonenko_C-int, ref:simha2023, ref:schultz2001, ref:Henkelman2000_climbing-neb, ref:wannier, ref:kohn-luttinger, ref:passler, ref:griffiths_rydberg, ref:si_exp-lattConst, ref:Si_dielConst, ref:alkauskas2012_huang-rhys, ref:telecom-region, ref:dornen85,ref:dornen86-prb,ref:dornen86,ref:dornen87,ref:dornen88,ref:davies89, ref:conzelmann1987, ref:sommerfeld, ref:turiansky_jpcm2024, ref:griffiths_hyd-1s-wf}

\bibliographystyle{apsrev4-2}
\bibliography{refs} 

\begin{thebibliography}{73}%
\makeatletter
\providecommand \@ifxundefined [1]{%
 \@ifx{#1\undefined}
}%
\providecommand \@ifnum [1]{%
 \ifnum #1\expandafter \@firstoftwo
 \else \expandafter \@secondoftwo
 \fi
}%
\providecommand \@ifx [1]{%
 \ifx #1\expandafter \@firstoftwo
 \else \expandafter \@secondoftwo
 \fi
}%
\providecommand \natexlab [1]{#1}%
\providecommand \enquote  [1]{``#1''}%
\providecommand \bibnamefont  [1]{#1}%
\providecommand \bibfnamefont [1]{#1}%
\providecommand \citenamefont [1]{#1}%
\providecommand \href@noop [0]{\@secondoftwo}%
\providecommand \href [0]{\begingroup \@sanitize@url \@href}%
\providecommand \@href[1]{\@@startlink{#1}\@@href}%
\providecommand \@@href[1]{\endgroup#1\@@endlink}%
\providecommand \@sanitize@url [0]{\catcode `\\12\catcode `\$12\catcode
  `\&12\catcode `\#12\catcode `\^12\catcode `\_12\catcode `\%12\relax}%
\providecommand \@@startlink[1]{}%
\providecommand \@@endlink[0]{}%
\providecommand \url  [0]{\begingroup\@sanitize@url \@url }%
\providecommand \@url [1]{\endgroup\@href {#1}{\urlprefix }}%
\providecommand \urlprefix  [0]{URL }%
\providecommand \Eprint [0]{\href }%
\providecommand \doibase [0]{https://doi.org/}%
\providecommand \selectlanguage [0]{\@gobble}%
\providecommand \bibinfo  [0]{\@secondoftwo}%
\providecommand \bibfield  [0]{\@secondoftwo}%
\providecommand \translation [1]{[#1]}%
\providecommand \BibitemOpen [0]{}%
\providecommand \bibitemStop [0]{}%
\providecommand \bibitemNoStop [0]{.\EOS\space}%
\providecommand \EOS [0]{\spacefactor3000\relax}%
\providecommand \BibitemShut  [1]{\csname bibitem#1\endcsname}%
\let\auto@bib@innerbib\@empty
\bibitem [{\citenamefont {Bradley}\ \emph {et~al.}(2019)\citenamefont
  {Bradley}, \citenamefont {Randall}, \citenamefont {Abobeih}, \citenamefont
  {Berrevoets}, \citenamefont {Degen}, \citenamefont {Bakker}, \citenamefont
  {Markham}, \citenamefont {Twitchen},\ and\ \citenamefont
  {Taminiau}}]{ref:bradley2019}%
  \BibitemOpen
  \bibfield  {author} {\bibinfo {author} {\bibfnamefont {C.~E.}\ \bibnamefont
  {Bradley}}, \bibinfo {author} {\bibfnamefont {J.}~\bibnamefont {Randall}},
  \bibinfo {author} {\bibfnamefont {M.~H.}\ \bibnamefont {Abobeih}}, \bibinfo
  {author} {\bibfnamefont {R.~C.}\ \bibnamefont {Berrevoets}}, \bibinfo
  {author} {\bibfnamefont {M.~J.}\ \bibnamefont {Degen}}, \bibinfo {author}
  {\bibfnamefont {M.~A.}\ \bibnamefont {Bakker}}, \bibinfo {author}
  {\bibfnamefont {M.}~\bibnamefont {Markham}}, \bibinfo {author} {\bibfnamefont
  {D.~J.}\ \bibnamefont {Twitchen}},\ and\ \bibinfo {author} {\bibfnamefont
  {T.~H.}\ \bibnamefont {Taminiau}},\ }\href
  {https://doi.org/10.1103/PhysRevX.9.031045} {\bibfield  {journal} {\bibinfo
  {journal} {Phys. Rev. X}\ }\textbf {\bibinfo {volume} {9}},\ \bibinfo {pages}
  {31045} (\bibinfo {year} {2019})}\BibitemShut {NoStop}%
\bibitem [{\citenamefont {Yan}\ \emph {et~al.}(2021)\citenamefont {Yan},
  \citenamefont {Gitt}, \citenamefont {Lin}, \citenamefont {Witt},
  \citenamefont {Abdolahi}, \citenamefont {Afifi}, \citenamefont {Azem},
  \citenamefont {Darcie}, \citenamefont {Wu}, \citenamefont {Awan},
  \citenamefont {Mitchell}, \citenamefont {Pfenning}, \citenamefont
  {Chrostowski},\ and\ \citenamefont {Young}}]{ref:yan2021}%
  \BibitemOpen
  \bibfield  {author} {\bibinfo {author} {\bibfnamefont {X.}~\bibnamefont
  {Yan}}, \bibinfo {author} {\bibfnamefont {S.}~\bibnamefont {Gitt}}, \bibinfo
  {author} {\bibfnamefont {B.}~\bibnamefont {Lin}}, \bibinfo {author}
  {\bibfnamefont {D.}~\bibnamefont {Witt}}, \bibinfo {author} {\bibfnamefont
  {M.}~\bibnamefont {Abdolahi}}, \bibinfo {author} {\bibfnamefont
  {A.}~\bibnamefont {Afifi}}, \bibinfo {author} {\bibfnamefont
  {A.}~\bibnamefont {Azem}}, \bibinfo {author} {\bibfnamefont {A.}~\bibnamefont
  {Darcie}}, \bibinfo {author} {\bibfnamefont {J.}~\bibnamefont {Wu}}, \bibinfo
  {author} {\bibfnamefont {K.}~\bibnamefont {Awan}}, \bibinfo {author}
  {\bibfnamefont {M.}~\bibnamefont {Mitchell}}, \bibinfo {author}
  {\bibfnamefont {A.}~\bibnamefont {Pfenning}}, \bibinfo {author}
  {\bibfnamefont {L.}~\bibnamefont {Chrostowski}},\ and\ \bibinfo {author}
  {\bibfnamefont {J.~F.}\ \bibnamefont {Young}},\ }\href
  {https://doi.org/10.1063/5.0049372} {\bibfield  {journal} {\bibinfo
  {journal} {APL Photonics}\ }\textbf {\bibinfo {volume} {6}},\ \bibinfo
  {pages} {070901} (\bibinfo {year} {2021})}\BibitemShut {NoStop}%
\bibitem [{\citenamefont {Simmons}(2024)}]{ref:simmons2024}%
  \BibitemOpen
  \bibfield  {author} {\bibinfo {author} {\bibfnamefont {S.}~\bibnamefont
  {Simmons}},\ }\href {https://doi.org/10.1103/PRXQuantum.5.010102} {\bibfield
  {journal} {\bibinfo  {journal} {PRX Quantum}\ }\textbf {\bibinfo {volume}
  {5}},\ \bibinfo {pages} {010102} (\bibinfo {year} {2024})}\BibitemShut
  {NoStop}%
\bibitem [{\citenamefont {Ruf}\ \emph {et~al.}(2021)\citenamefont {Ruf},
  \citenamefont {Wan}, \citenamefont {Choi}, \citenamefont {Englund},\ and\
  \citenamefont {Hanson}}]{ref:ruf2021}%
  \BibitemOpen
  \bibfield  {author} {\bibinfo {author} {\bibfnamefont {M.}~\bibnamefont
  {Ruf}}, \bibinfo {author} {\bibfnamefont {N.~H.}\ \bibnamefont {Wan}},
  \bibinfo {author} {\bibfnamefont {H.}~\bibnamefont {Choi}}, \bibinfo {author}
  {\bibfnamefont {D.}~\bibnamefont {Englund}},\ and\ \bibinfo {author}
  {\bibfnamefont {R.}~\bibnamefont {Hanson}},\ }\href
  {https://doi.org/10.1063/5.0056534} {\bibfield  {journal} {\bibinfo
  {journal} {J. Appl. Phys.}\ }\textbf {\bibinfo {volume} {130}},\ \bibinfo
  {pages} {070901} (\bibinfo {year} {2021})}\BibitemShut {NoStop}%
\bibitem [{\citenamefont {Northup}\ and\ \citenamefont
  {Blatt}(2014)}]{ref:northup2014}%
  \BibitemOpen
  \bibfield  {author} {\bibinfo {author} {\bibfnamefont {T.~E.}\ \bibnamefont
  {Northup}}\ and\ \bibinfo {author} {\bibfnamefont {R.}~\bibnamefont
  {Blatt}},\ }\href {https://doi.org/10.1038/nphoton.2014.53} {\bibfield
  {journal} {\bibinfo  {journal} {Nat. Photonics}\ }\textbf {\bibinfo {volume}
  {8}},\ \bibinfo {pages} {356} (\bibinfo {year} {2014})}\BibitemShut {NoStop}%
\bibitem [{\citenamefont {Turiansky}\ \emph
  {et~al.}(2024{\natexlab{a}})\citenamefont {Turiansky}, \citenamefont {Parto},
  \citenamefont {Moody},\ and\ \citenamefont {{Van de
  Walle}}}]{ref:turiansky_rational}%
  \BibitemOpen
  \bibfield  {author} {\bibinfo {author} {\bibfnamefont {M.~E.}\ \bibnamefont
  {Turiansky}}, \bibinfo {author} {\bibfnamefont {K.}~\bibnamefont {Parto}},
  \bibinfo {author} {\bibfnamefont {G.}~\bibnamefont {Moody}},\ and\ \bibinfo
  {author} {\bibfnamefont {C.~G.}\ \bibnamefont {{Van de Walle}}},\ }\href
  {https://doi.org/10.1063/5.0203366} {\bibfield  {journal} {\bibinfo
  {journal} {APL Photonics}\ }\textbf {\bibinfo {volume} {9}},\ \bibinfo
  {pages} {066117} (\bibinfo {year} {2024}{\natexlab{a}})}\BibitemShut
  {NoStop}%
\bibitem [{\citenamefont {Weber}\ \emph {et~al.}(2010)\citenamefont {Weber},
  \citenamefont {Koehl}, \citenamefont {Varley}, \citenamefont {Janotti},
  \citenamefont {Buckley}, \citenamefont {{Van de Walle}},\ and\ \citenamefont
  {Awschalom}}]{ref:weber2010}%
  \BibitemOpen
  \bibfield  {author} {\bibinfo {author} {\bibfnamefont {J.~R.}\ \bibnamefont
  {Weber}}, \bibinfo {author} {\bibfnamefont {W.~F.}\ \bibnamefont {Koehl}},
  \bibinfo {author} {\bibfnamefont {J.~B.}\ \bibnamefont {Varley}}, \bibinfo
  {author} {\bibfnamefont {A.}~\bibnamefont {Janotti}}, \bibinfo {author}
  {\bibfnamefont {B.~B.}\ \bibnamefont {Buckley}}, \bibinfo {author}
  {\bibfnamefont {C.~G.}\ \bibnamefont {{Van de Walle}}},\ and\ \bibinfo
  {author} {\bibfnamefont {D.~D.}\ \bibnamefont {Awschalom}},\ }\href
  {https://doi.org/10.1073/pnas.1003052107} {\bibfield  {journal} {\bibinfo
  {journal} {Proc. Natl. Acad. Sci.}\ }\textbf {\bibinfo {volume} {107}},\
  \bibinfo {pages} {8513} (\bibinfo {year} {2010})}\BibitemShut {NoStop}%
\bibitem [{\citenamefont {Davies}(1981)}]{ref:davies81}%
  \BibitemOpen
  \bibfield  {author} {\bibinfo {author} {\bibfnamefont {G.}~\bibnamefont
  {Davies}},\ }\href {https://doi.org/10.1088/0034-4885/44/7/003} {\bibfield
  {journal} {\bibinfo  {journal} {Reports Prog. Phys.}\ }\textbf {\bibinfo
  {volume} {44}},\ \bibinfo {pages} {787} (\bibinfo {year} {1981})}\BibitemShut
  {NoStop}%
\bibitem [{\citenamefont {Bradac}\ \emph {et~al.}(2019)\citenamefont {Bradac},
  \citenamefont {Gao}, \citenamefont {Forneris}, \citenamefont {Trusheim},\
  and\ \citenamefont {Aharonovich}}]{ref:bradac2019}%
  \BibitemOpen
  \bibfield  {author} {\bibinfo {author} {\bibfnamefont {C.}~\bibnamefont
  {Bradac}}, \bibinfo {author} {\bibfnamefont {W.}~\bibnamefont {Gao}},
  \bibinfo {author} {\bibfnamefont {J.}~\bibnamefont {Forneris}}, \bibinfo
  {author} {\bibfnamefont {M.~E.}\ \bibnamefont {Trusheim}},\ and\ \bibinfo
  {author} {\bibfnamefont {I.}~\bibnamefont {Aharonovich}},\ }\href
  {https://doi.org/10.1038/s41467-019-13332-w} {\bibfield  {journal} {\bibinfo
  {journal} {Nat. Commun.}\ }\textbf {\bibinfo {volume} {10}},\ \bibinfo
  {pages} {5625} (\bibinfo {year} {2019})}\BibitemShut {NoStop}%
\bibitem [{\citenamefont {Turiansky}\ and\ \citenamefont {{Van de
  Walle}}(2023)}]{ref:turiansky-cubic-bn}%
  \BibitemOpen
  \bibfield  {author} {\bibinfo {author} {\bibfnamefont {M.~E.}\ \bibnamefont
  {Turiansky}}\ and\ \bibinfo {author} {\bibfnamefont {C.~G.}\ \bibnamefont
  {{Van de Walle}}},\ }\href {https://doi.org/10.1103/PhysRevB.108.L041102}
  {\bibfield  {journal} {\bibinfo  {journal} {Phys. Rev. B}\ }\textbf {\bibinfo
  {volume} {108}},\ \bibinfo {pages} {L041102} (\bibinfo {year}
  {2023})}\BibitemShut {NoStop}%
\bibitem [{\citenamefont {Bergeron}\ \emph {et~al.}(2020)\citenamefont
  {Bergeron}, \citenamefont {Chartrand}, \citenamefont {Kurkjian},
  \citenamefont {Morse}, \citenamefont {Riemann}, \citenamefont {Abrosimov},
  \citenamefont {Becker}, \citenamefont {Pohl}, \citenamefont {Thewalt},\ and\
  \citenamefont {Simmons}}]{ref:bergeron}%
  \BibitemOpen
  \bibfield  {author} {\bibinfo {author} {\bibfnamefont {L.}~\bibnamefont
  {Bergeron}}, \bibinfo {author} {\bibfnamefont {C.}~\bibnamefont {Chartrand}},
  \bibinfo {author} {\bibfnamefont {A.~T.~K.}\ \bibnamefont {Kurkjian}},
  \bibinfo {author} {\bibfnamefont {K.~J.}\ \bibnamefont {Morse}}, \bibinfo
  {author} {\bibfnamefont {H.}~\bibnamefont {Riemann}}, \bibinfo {author}
  {\bibfnamefont {N.~V.}\ \bibnamefont {Abrosimov}}, \bibinfo {author}
  {\bibfnamefont {P.}~\bibnamefont {Becker}}, \bibinfo {author} {\bibfnamefont
  {H.-J.}\ \bibnamefont {Pohl}}, \bibinfo {author} {\bibfnamefont {M.~L.~W.}\
  \bibnamefont {Thewalt}},\ and\ \bibinfo {author} {\bibfnamefont
  {S.}~\bibnamefont {Simmons}},\ }\href
  {https://doi.org/10.1103/PRXQuantum.1.020301} {\bibfield  {journal} {\bibinfo
   {journal} {PRX Quantum}\ }\textbf {\bibinfo {volume} {1}},\ \bibinfo {pages}
  {020301} (\bibinfo {year} {2020})}\BibitemShut {NoStop}%
\bibitem [{\citenamefont {Dhaliah}\ \emph {et~al.}(2022)\citenamefont
  {Dhaliah}, \citenamefont {Xiong}, \citenamefont {Sipahigil}, \citenamefont
  {Griffin},\ and\ \citenamefont {Hautier}}]{ref:dhaliah}%
  \BibitemOpen
  \bibfield  {author} {\bibinfo {author} {\bibfnamefont {D.}~\bibnamefont
  {Dhaliah}}, \bibinfo {author} {\bibfnamefont {Y.}~\bibnamefont {Xiong}},
  \bibinfo {author} {\bibfnamefont {A.}~\bibnamefont {Sipahigil}}, \bibinfo
  {author} {\bibfnamefont {S.~M.}\ \bibnamefont {Griffin}},\ and\ \bibinfo
  {author} {\bibfnamefont {G.}~\bibnamefont {Hautier}},\ }\href
  {https://doi.org/10.1103/PhysRevMaterials.6.L053201} {\bibfield  {journal}
  {\bibinfo  {journal} {Phys. Rev. Mater.}\ }\textbf {\bibinfo {volume} {6}},\
  \bibinfo {pages} {L053201} (\bibinfo {year} {2022})}\BibitemShut {NoStop}%
\bibitem [{\citenamefont {MacQuarrie}\ \emph {et~al.}(2021)\citenamefont
  {MacQuarrie}, \citenamefont {Chartrand}, \citenamefont {Higginbottom},
  \citenamefont {Morse}, \citenamefont {Karasyuk}, \citenamefont {Roorda},\
  and\ \citenamefont {Simmons}}]{ref:macquarrie}%
  \BibitemOpen
  \bibfield  {author} {\bibinfo {author} {\bibfnamefont {E.~R.}\ \bibnamefont
  {MacQuarrie}}, \bibinfo {author} {\bibfnamefont {C.}~\bibnamefont
  {Chartrand}}, \bibinfo {author} {\bibfnamefont {D.~B.}\ \bibnamefont
  {Higginbottom}}, \bibinfo {author} {\bibfnamefont {K.~J.}\ \bibnamefont
  {Morse}}, \bibinfo {author} {\bibfnamefont {V.~A.}\ \bibnamefont {Karasyuk}},
  \bibinfo {author} {\bibfnamefont {S.}~\bibnamefont {Roorda}},\ and\ \bibinfo
  {author} {\bibfnamefont {S.}~\bibnamefont {Simmons}},\ }\href
  {https://doi.org/10.1088/1367-2630/ac291f} {\bibfield  {journal} {\bibinfo
  {journal} {New J. Phys.}\ }\textbf {\bibinfo {volume} {23}},\ \bibinfo
  {pages} {103008} (\bibinfo {year} {2021})}\BibitemShut {NoStop}%
\bibitem [{\citenamefont {Higginbottom}\ \emph {et~al.}(2022)\citenamefont
  {Higginbottom}, \citenamefont {Kurkjian}, \citenamefont {Chartrand},
  \citenamefont {Kazemi}, \citenamefont {Brunelle}, \citenamefont {MacQuarrie},
  \citenamefont {Klein}, \citenamefont {Lee-Hone}, \citenamefont {Stacho},
  \citenamefont {Ruether}, \citenamefont {Bowness}, \citenamefont {Bergeron},
  \citenamefont {DeAbreu}, \citenamefont {Harrigan}, \citenamefont
  {Kanaganayagam}, \citenamefont {Marsden}, \citenamefont {Richards},
  \citenamefont {Stott}, \citenamefont {Roorda}, \citenamefont {Morse},
  \citenamefont {Thewalt},\ and\ \citenamefont {Simmons}}]{ref:higginbottom22}%
  \BibitemOpen
  \bibfield  {author} {\bibinfo {author} {\bibfnamefont {D.~B.}\ \bibnamefont
  {Higginbottom}}, \bibinfo {author} {\bibfnamefont {A.~T.}\ \bibnamefont
  {Kurkjian}}, \bibinfo {author} {\bibfnamefont {C.}~\bibnamefont {Chartrand}},
  \bibinfo {author} {\bibfnamefont {M.}~\bibnamefont {Kazemi}}, \bibinfo
  {author} {\bibfnamefont {N.~A.}\ \bibnamefont {Brunelle}}, \bibinfo {author}
  {\bibfnamefont {E.~R.}\ \bibnamefont {MacQuarrie}}, \bibinfo {author}
  {\bibfnamefont {J.~R.}\ \bibnamefont {Klein}}, \bibinfo {author}
  {\bibfnamefont {N.~R.}\ \bibnamefont {Lee-Hone}}, \bibinfo {author}
  {\bibfnamefont {J.}~\bibnamefont {Stacho}}, \bibinfo {author} {\bibfnamefont
  {M.}~\bibnamefont {Ruether}}, \bibinfo {author} {\bibfnamefont
  {C.}~\bibnamefont {Bowness}}, \bibinfo {author} {\bibfnamefont
  {L.}~\bibnamefont {Bergeron}}, \bibinfo {author} {\bibfnamefont
  {A.}~\bibnamefont {DeAbreu}}, \bibinfo {author} {\bibfnamefont {S.~R.}\
  \bibnamefont {Harrigan}}, \bibinfo {author} {\bibfnamefont {J.}~\bibnamefont
  {Kanaganayagam}}, \bibinfo {author} {\bibfnamefont {D.~W.}\ \bibnamefont
  {Marsden}}, \bibinfo {author} {\bibfnamefont {T.~S.}\ \bibnamefont
  {Richards}}, \bibinfo {author} {\bibfnamefont {L.~A.}\ \bibnamefont {Stott}},
  \bibinfo {author} {\bibfnamefont {S.}~\bibnamefont {Roorda}}, \bibinfo
  {author} {\bibfnamefont {K.~J.}\ \bibnamefont {Morse}}, \bibinfo {author}
  {\bibfnamefont {M.~L.}\ \bibnamefont {Thewalt}},\ and\ \bibinfo {author}
  {\bibfnamefont {S.}~\bibnamefont {Simmons}},\ }\href
  {https://doi.org/10.1038/s41586-022-04821-y} {\bibfield  {journal} {\bibinfo
  {journal} {Nature}\ }\textbf {\bibinfo {volume} {607}},\ \bibinfo {pages}
  {266} (\bibinfo {year} {2022})}\BibitemShut {NoStop}%
\bibitem [{\citenamefont {Higginbottom}\ \emph {et~al.}(2023)\citenamefont
  {Higginbottom}, \citenamefont {Asadi}, \citenamefont {Chartrand},
  \citenamefont {Ji}, \citenamefont {Bergeron}, \citenamefont {Thewalt},
  \citenamefont {Simon},\ and\ \citenamefont {Simmons}}]{ref:higginbottom23}%
  \BibitemOpen
  \bibfield  {author} {\bibinfo {author} {\bibfnamefont {D.~B.}\ \bibnamefont
  {Higginbottom}}, \bibinfo {author} {\bibfnamefont {F.~K.}\ \bibnamefont
  {Asadi}}, \bibinfo {author} {\bibfnamefont {C.}~\bibnamefont {Chartrand}},
  \bibinfo {author} {\bibfnamefont {J.~W.}\ \bibnamefont {Ji}}, \bibinfo
  {author} {\bibfnamefont {L.}~\bibnamefont {Bergeron}}, \bibinfo {author}
  {\bibfnamefont {M.~L.}\ \bibnamefont {Thewalt}}, \bibinfo {author}
  {\bibfnamefont {C.}~\bibnamefont {Simon}},\ and\ \bibinfo {author}
  {\bibfnamefont {S.}~\bibnamefont {Simmons}},\ }\href
  {https://doi.org/10.1103/PRXQuantum.4.020308} {\bibfield  {journal} {\bibinfo
   {journal} {PRX Quantum}\ }\textbf {\bibinfo {volume} {4}},\ \bibinfo {pages}
  {020308} (\bibinfo {year} {2023})}\BibitemShut {NoStop}%
\bibitem [{\citenamefont {DeAbreu}\ \emph {et~al.}(2023)\citenamefont
  {DeAbreu}, \citenamefont {Bowness}, \citenamefont {Alizadeh}, \citenamefont
  {Chartrand}, \citenamefont {Brunelle}, \citenamefont {MacQuarrie},
  \citenamefont {Lee-Hone}, \citenamefont {Ruether}, \citenamefont {Kazemi},
  \citenamefont {Kurkjian}, \citenamefont {Roorda}, \citenamefont {Abrosimov},
  \citenamefont {Pohl}, \citenamefont {Thewalt}, \citenamefont {Higginbottom},\
  and\ \citenamefont {Simmons}}]{ref:deabreu}%
  \BibitemOpen
  \bibfield  {author} {\bibinfo {author} {\bibfnamefont {A.}~\bibnamefont
  {DeAbreu}}, \bibinfo {author} {\bibfnamefont {C.}~\bibnamefont {Bowness}},
  \bibinfo {author} {\bibfnamefont {A.}~\bibnamefont {Alizadeh}}, \bibinfo
  {author} {\bibfnamefont {C.}~\bibnamefont {Chartrand}}, \bibinfo {author}
  {\bibfnamefont {N.~A.}\ \bibnamefont {Brunelle}}, \bibinfo {author}
  {\bibfnamefont {E.~R.}\ \bibnamefont {MacQuarrie}}, \bibinfo {author}
  {\bibfnamefont {N.~R.}\ \bibnamefont {Lee-Hone}}, \bibinfo {author}
  {\bibfnamefont {M.}~\bibnamefont {Ruether}}, \bibinfo {author} {\bibfnamefont
  {M.}~\bibnamefont {Kazemi}}, \bibinfo {author} {\bibfnamefont {A.~T.~K.}\
  \bibnamefont {Kurkjian}}, \bibinfo {author} {\bibfnamefont {S.}~\bibnamefont
  {Roorda}}, \bibinfo {author} {\bibfnamefont {N.~V.}\ \bibnamefont
  {Abrosimov}}, \bibinfo {author} {\bibfnamefont {H.-J.}\ \bibnamefont {Pohl}},
  \bibinfo {author} {\bibfnamefont {M.~L.~W.}\ \bibnamefont {Thewalt}},
  \bibinfo {author} {\bibfnamefont {D.~B.}\ \bibnamefont {Higginbottom}},\ and\
  \bibinfo {author} {\bibfnamefont {S.}~\bibnamefont {Simmons}},\ }\href
  {https://doi.org/10.1364/oe.482008} {\bibfield  {journal} {\bibinfo
  {journal} {Opt. Express}\ }\textbf {\bibinfo {volume} {31}},\ \bibinfo
  {pages} {15045} (\bibinfo {year} {2023})}\BibitemShut {NoStop}%
\bibitem [{\citenamefont {Islam}\ \emph {et~al.}(2024)\citenamefont {Islam},
  \citenamefont {Lee}, \citenamefont {Harper}, \citenamefont {Rahaman},
  \citenamefont {Zhao}, \citenamefont {Vij},\ and\ \citenamefont
  {Waks}}]{ref:islam}%
  \BibitemOpen
  \bibfield  {author} {\bibinfo {author} {\bibfnamefont {F.}~\bibnamefont
  {Islam}}, \bibinfo {author} {\bibfnamefont {C.-M.}\ \bibnamefont {Lee}},
  \bibinfo {author} {\bibfnamefont {S.}~\bibnamefont {Harper}}, \bibinfo
  {author} {\bibfnamefont {M.~H.}\ \bibnamefont {Rahaman}}, \bibinfo {author}
  {\bibfnamefont {Y.}~\bibnamefont {Zhao}}, \bibinfo {author} {\bibfnamefont
  {N.~K.}\ \bibnamefont {Vij}},\ and\ \bibinfo {author} {\bibfnamefont
  {E.}~\bibnamefont {Waks}},\ }\href
  {https://doi.org/10.1021/acs.nanolett.3c04056} {\bibfield  {journal}
  {\bibinfo  {journal} {Nano Lett.}\ }\textbf {\bibinfo {volume} {24}},\
  \bibinfo {pages} {319} (\bibinfo {year} {2024})}\BibitemShut {NoStop}%
\bibitem [{\citenamefont {Bl\"ochl}(1994)}]{ref:paw}%
  \BibitemOpen
  \bibfield  {author} {\bibinfo {author} {\bibfnamefont {P.~E.}\ \bibnamefont
  {Bl\"ochl}},\ }\href {https://doi.org/10.1103/PhysRevB.50.17953} {\bibfield
  {journal} {\bibinfo  {journal} {Phys. Rev. B}\ }\textbf {\bibinfo {volume}
  {50}},\ \bibinfo {pages} {17953} (\bibinfo {year} {1994})}\BibitemShut
  {NoStop}%
\bibitem [{\citenamefont {Kresse}\ and\ \citenamefont
  {Joubert}(1999)}]{ref:vasp-paw}%
  \BibitemOpen
  \bibfield  {author} {\bibinfo {author} {\bibfnamefont {G.}~\bibnamefont
  {Kresse}}\ and\ \bibinfo {author} {\bibfnamefont {D.}~\bibnamefont
  {Joubert}},\ }\href {https://doi.org/10.1103/PhysRevB.59.1758} {\bibfield
  {journal} {\bibinfo  {journal} {Phys. Rev. B}\ }\textbf {\bibinfo {volume}
  {59}},\ \bibinfo {pages} {1758} (\bibinfo {year} {1999})}\BibitemShut
  {NoStop}%
\bibitem [{\citenamefont {Kresse}\ and\ \citenamefont
  {Furthmüller}(1996)}]{ref:vasp}%
  \BibitemOpen
  \bibfield  {author} {\bibinfo {author} {\bibfnamefont {G.}~\bibnamefont
  {Kresse}}\ and\ \bibinfo {author} {\bibfnamefont {J.}~\bibnamefont
  {Furthmüller}},\ }\href
  {https://doi.org/https://doi.org/10.1016/0927-0256(96)00008-0} {\bibfield
  {journal} {\bibinfo  {journal} {Comput. Mater. Sci.}\ }\textbf {\bibinfo
  {volume} {6}},\ \bibinfo {pages} {15} (\bibinfo {year} {1996})}\BibitemShut
  {NoStop}%
\bibitem [{\citenamefont {Kresse}\ and\ \citenamefont
  {Furthm\"uller}(1996)}]{ref:vasp2}%
  \BibitemOpen
  \bibfield  {author} {\bibinfo {author} {\bibfnamefont {G.}~\bibnamefont
  {Kresse}}\ and\ \bibinfo {author} {\bibfnamefont {J.}~\bibnamefont
  {Furthm\"uller}},\ }\href {https://doi.org/10.1103/PhysRevB.54.11169}
  {\bibfield  {journal} {\bibinfo  {journal} {Phys. Rev. B}\ }\textbf {\bibinfo
  {volume} {54}},\ \bibinfo {pages} {11169} (\bibinfo {year}
  {1996})}\BibitemShut {NoStop}%
\bibitem [{\citenamefont {Heyd}\ \emph {et~al.}(2003)\citenamefont {Heyd},
  \citenamefont {Scuseria},\ and\ \citenamefont {Ernzerhof}}]{ref:HSE03}%
  \BibitemOpen
  \bibfield  {author} {\bibinfo {author} {\bibfnamefont {J.}~\bibnamefont
  {Heyd}}, \bibinfo {author} {\bibfnamefont {G.~E.}\ \bibnamefont {Scuseria}},\
  and\ \bibinfo {author} {\bibfnamefont {M.}~\bibnamefont {Ernzerhof}},\ }\href
  {https://doi.org/10.1063/1.1564060} {\bibfield  {journal} {\bibinfo
  {journal} {J. Chem. Phys.}\ }\textbf {\bibinfo {volume} {118}},\ \bibinfo
  {pages} {8207} (\bibinfo {year} {2003})}\BibitemShut {NoStop}%
\bibitem [{\citenamefont {Heyd}\ \emph {et~al.}(2006)\citenamefont {Heyd},
  \citenamefont {Scuseria},\ and\ \citenamefont {Ernzerhof}}]{ref:HSE06}%
  \BibitemOpen
  \bibfield  {author} {\bibinfo {author} {\bibfnamefont {J.}~\bibnamefont
  {Heyd}}, \bibinfo {author} {\bibfnamefont {G.~E.}\ \bibnamefont {Scuseria}},\
  and\ \bibinfo {author} {\bibfnamefont {M.}~\bibnamefont {Ernzerhof}},\ }\href
  {https://doi.org/10.1063/1.2204597} {\bibfield  {journal} {\bibinfo
  {journal} {J. Chem. Phys.}\ }\textbf {\bibinfo {volume} {124}},\ \bibinfo
  {pages} {219906} (\bibinfo {year} {2006})}\BibitemShut {NoStop}%
\bibitem [{\citenamefont {Perdew}\ \emph {et~al.}(1996)\citenamefont {Perdew},
  \citenamefont {Ernzerhof},\ and\ \citenamefont
  {Burke}}]{ref:perdew1996_pbe0}%
  \BibitemOpen
  \bibfield  {author} {\bibinfo {author} {\bibfnamefont {J.~P.}\ \bibnamefont
  {Perdew}}, \bibinfo {author} {\bibfnamefont {M.}~\bibnamefont {Ernzerhof}},\
  and\ \bibinfo {author} {\bibfnamefont {K.}~\bibnamefont {Burke}},\ }\href
  {https://doi.org/10.1063/1.472933} {\bibfield  {journal} {\bibinfo  {journal}
  {J. Chem. Phys.}\ }\textbf {\bibinfo {volume} {105}},\ \bibinfo {pages}
  {9982} (\bibinfo {year} {1996})}\BibitemShut {NoStop}%
\bibitem [{\citenamefont {Ernzerhof}\ and\ \citenamefont
  {Scuseria}(1999)}]{ref:ernzerhof1999_pbe0}%
  \BibitemOpen
  \bibfield  {author} {\bibinfo {author} {\bibfnamefont {M.}~\bibnamefont
  {Ernzerhof}}\ and\ \bibinfo {author} {\bibfnamefont {G.~E.}\ \bibnamefont
  {Scuseria}},\ }\href {https://doi.org/10.1063/1.478401} {\bibfield  {journal}
  {\bibinfo  {journal} {J. Chem. Phys.}\ }\textbf {\bibinfo {volume} {110}},\
  \bibinfo {pages} {5029} (\bibinfo {year} {1999})}\BibitemShut {NoStop}%
\bibitem [{\citenamefont {Carlo}\ and\ \citenamefont
  {Barone}(1999)}]{ref:adamo1999_pbe0}%
  \BibitemOpen
  \bibfield  {author} {\bibinfo {author} {\bibnamefont {Carlo}}\ and\ \bibinfo
  {author} {\bibfnamefont {V.}~\bibnamefont {Barone}},\ }\href
  {https://doi.org/10.1063/1.478522} {\bibfield  {journal} {\bibinfo  {journal}
  {J. Chem. Phys.}\ }\textbf {\bibinfo {volume} {110}},\ \bibinfo {pages}
  {6158} (\bibinfo {year} {1999})}\BibitemShut {NoStop}%
\bibitem [{\citenamefont {Shur}(1996)}]{ref:si_exp-lattConst}%
  \BibitemOpen
  \bibfield  {author} {\bibinfo {author} {\bibfnamefont {M.~S.}\ \bibnamefont
  {Shur}},\ }\href@noop {} {\emph {\bibinfo {title} {{Handbook Series on
  Semiconductor Parameters}}}},\ Vol.~\bibinfo {volume} {1}\ (\bibinfo
  {publisher} {World Scientific},\ \bibinfo {year} {1996})\BibitemShut
  {NoStop}%
\bibitem [{\citenamefont {Bludau}\ \emph {et~al.}(1974)\citenamefont {Bludau},
  \citenamefont {Onton},\ and\ \citenamefont {Heinke}}]{ref:si_exp_gap}%
  \BibitemOpen
  \bibfield  {author} {\bibinfo {author} {\bibfnamefont {W.}~\bibnamefont
  {Bludau}}, \bibinfo {author} {\bibfnamefont {A.}~\bibnamefont {Onton}},\ and\
  \bibinfo {author} {\bibfnamefont {W.}~\bibnamefont {Heinke}},\ }\href
  {https://doi.org/10.1063/1.1663501} {\bibfield  {journal} {\bibinfo
  {journal} {J. Appl. Phys.}\ }\textbf {\bibinfo {volume} {45}},\ \bibinfo
  {pages} {1846} (\bibinfo {year} {1974})}\BibitemShut {NoStop}%
\bibitem [{\citenamefont {Platonenko}\ \emph {et~al.}(2019)\citenamefont
  {Platonenko}, \citenamefont {Gentile}, \citenamefont {Maul}, \citenamefont
  {Pascale}, \citenamefont {Kotomin},\ and\ \citenamefont
  {Dovesi}}]{ref:platonenko_N-int}%
  \BibitemOpen
  \bibfield  {author} {\bibinfo {author} {\bibfnamefont {A.}~\bibnamefont
  {Platonenko}}, \bibinfo {author} {\bibfnamefont {F.~S.}\ \bibnamefont
  {Gentile}}, \bibinfo {author} {\bibfnamefont {J.}~\bibnamefont {Maul}},
  \bibinfo {author} {\bibfnamefont {F.}~\bibnamefont {Pascale}}, \bibinfo
  {author} {\bibfnamefont {E.~A.}\ \bibnamefont {Kotomin}},\ and\ \bibinfo
  {author} {\bibfnamefont {R.}~\bibnamefont {Dovesi}},\ }\href
  {https://doi.org/10.1016/j.mtcomm.2019.100616} {\bibfield  {journal}
  {\bibinfo  {journal} {Mater. Today Commun.}\ }\textbf {\bibinfo {volume}
  {21}},\ \bibinfo {pages} {100616} (\bibinfo {year} {2019})}\BibitemShut
  {NoStop}%
\bibitem [{\citenamefont {Kuganathan}\ \emph {et~al.}(2023)\citenamefont
  {Kuganathan}, \citenamefont {Christopoulos}, \citenamefont {Papadopoulou},
  \citenamefont {Sgourou}, \citenamefont {Chroneos},\ and\ \citenamefont
  {Londos}}]{ref:kuganathan2023}%
  \BibitemOpen
  \bibfield  {author} {\bibinfo {author} {\bibfnamefont {N.}~\bibnamefont
  {Kuganathan}}, \bibinfo {author} {\bibfnamefont {S.-R.~G.}\ \bibnamefont
  {Christopoulos}}, \bibinfo {author} {\bibfnamefont {K.}~\bibnamefont
  {Papadopoulou}}, \bibinfo {author} {\bibfnamefont {E.~N.}\ \bibnamefont
  {Sgourou}}, \bibinfo {author} {\bibfnamefont {A.}~\bibnamefont {Chroneos}},\
  and\ \bibinfo {author} {\bibfnamefont {C.~A.}\ \bibnamefont {Londos}},\
  }\href {https://doi.org/10.1142/S0217984923501543} {\bibfield  {journal}
  {\bibinfo  {journal} {Mod. Phys. Lett. B}\ }\textbf {\bibinfo {volume}
  {37}},\ \bibinfo {pages} {2350154} (\bibinfo {year} {2023})}\BibitemShut
  {NoStop}%
\bibitem [{\citenamefont {Sgourou}\ \emph {et~al.}(2024)\citenamefont
  {Sgourou}, \citenamefont {Sarlis}, \citenamefont {Chroneos},\ and\
  \citenamefont {Londos}}]{ref:sgourou2024}%
  \BibitemOpen
  \bibfield  {author} {\bibinfo {author} {\bibfnamefont {E.~N.}\ \bibnamefont
  {Sgourou}}, \bibinfo {author} {\bibfnamefont {N.}~\bibnamefont {Sarlis}},
  \bibinfo {author} {\bibfnamefont {A.}~\bibnamefont {Chroneos}},\ and\
  \bibinfo {author} {\bibfnamefont {C.~A.}\ \bibnamefont {Londos}},\ }\href
  {https://doi.org/10.3390/app14041631} {\bibfield  {journal} {\bibinfo
  {journal} {Appl. Sci.}\ }\textbf {\bibinfo {volume} {14}},\ \bibinfo {pages}
  {1631} (\bibinfo {year} {2024})}\BibitemShut {NoStop}%
\bibitem [{\citenamefont {Freysoldt}\ \emph {et~al.}(2014)\citenamefont
  {Freysoldt}, \citenamefont {Grabowski}, \citenamefont {Hickel}, \citenamefont
  {Neugebauer}, \citenamefont {Kresse}, \citenamefont {Janotti},\ and\
  \citenamefont {{Van de Walle}}}]{ref:vdw-review}%
  \BibitemOpen
  \bibfield  {author} {\bibinfo {author} {\bibfnamefont {C.}~\bibnamefont
  {Freysoldt}}, \bibinfo {author} {\bibfnamefont {B.}~\bibnamefont
  {Grabowski}}, \bibinfo {author} {\bibfnamefont {T.}~\bibnamefont {Hickel}},
  \bibinfo {author} {\bibfnamefont {J.}~\bibnamefont {Neugebauer}}, \bibinfo
  {author} {\bibfnamefont {G.}~\bibnamefont {Kresse}}, \bibinfo {author}
  {\bibfnamefont {A.}~\bibnamefont {Janotti}},\ and\ \bibinfo {author}
  {\bibfnamefont {C.~G.}\ \bibnamefont {{Van de Walle}}},\ }\href
  {https://doi.org/10.1103/RevModPhys.86.253} {\bibfield  {journal} {\bibinfo
  {journal} {Rev. Mod. Phys.}\ }\textbf {\bibinfo {volume} {86}},\ \bibinfo
  {pages} {253} (\bibinfo {year} {2014})}\BibitemShut {NoStop}%
\bibitem [{\citenamefont {Freysoldt}\ \emph {et~al.}(2009)\citenamefont
  {Freysoldt}, \citenamefont {Neugebauer},\ and\ \citenamefont {{Van de
  Walle}}}]{ref:fnv}%
  \BibitemOpen
  \bibfield  {author} {\bibinfo {author} {\bibfnamefont {C.}~\bibnamefont
  {Freysoldt}}, \bibinfo {author} {\bibfnamefont {J.}~\bibnamefont
  {Neugebauer}},\ and\ \bibinfo {author} {\bibfnamefont {C.~G.}\ \bibnamefont
  {{Van de Walle}}},\ }\href {https://doi.org/10.1103/PhysRevLett.102.016402}
  {\bibfield  {journal} {\bibinfo  {journal} {Phys. Rev. Lett.}\ }\textbf
  {\bibinfo {volume} {102}},\ \bibinfo {pages} {016402} (\bibinfo {year}
  {2009})}\BibitemShut {NoStop}%
\bibitem [{ref()}]{ref:supplemental}%
  \BibitemOpen
  \href@noop {} {}\bibinfo {note} {{See Supplemental Material at [URL will be
  inserted by publisher] for Kohn-Sham states of non-neutral charge states of T
  and CN centers, other defects considered, migration barrier calculations,
  bound exciton supercell-size dependence, Huang-Rhys factor calculations, full
  vs. single-shot PBE0, all zero-phonon lines, radiative lifetime calculations,
  and relation of radiative lifetime to the radiative capture formalism, which
  includes Refs.~\cite{ref:schultz2001, ref:wannier, ref:kohn-luttinger,
  ref:passler, ref:griffiths_rydberg, ref:Si_dielConst, ref:conzelmann1987,
  ref:sommerfeld, ref:turiansky_jpcm2024,
  ref:griffiths_hyd-1s-wf}.}}\BibitemShut {Stop}%
\bibitem [{\citenamefont {Tersoff}(1990)}]{ref:tersoff1990}%
  \BibitemOpen
  \bibfield  {author} {\bibinfo {author} {\bibfnamefont {J.}~\bibnamefont
  {Tersoff}},\ }\href {https://doi.org/10.1103/PhysRevLett.64.1757} {\bibfield
  {journal} {\bibinfo  {journal} {Phys. Rev. Lett.}\ }\textbf {\bibinfo
  {volume} {64}},\ \bibinfo {pages} {1757} (\bibinfo {year}
  {1990})}\BibitemShut {NoStop}%
\bibitem [{\citenamefont {Gentile}\ \emph {et~al.}(2020)\citenamefont
  {Gentile}, \citenamefont {Platonenko}, \citenamefont {El-Kelany},
  \citenamefont {R{\'{e}}rat}, \citenamefont {D'Arco},\ and\ \citenamefont
  {Dovesi}}]{ref:platonenko_C-sub}%
  \BibitemOpen
  \bibfield  {author} {\bibinfo {author} {\bibfnamefont {F.~S.}\ \bibnamefont
  {Gentile}}, \bibinfo {author} {\bibfnamefont {A.}~\bibnamefont {Platonenko}},
  \bibinfo {author} {\bibfnamefont {K.~E.}\ \bibnamefont {El-Kelany}}, \bibinfo
  {author} {\bibfnamefont {M.}~\bibnamefont {R{\'{e}}rat}}, \bibinfo {author}
  {\bibfnamefont {P.}~\bibnamefont {D'Arco}},\ and\ \bibinfo {author}
  {\bibfnamefont {R.}~\bibnamefont {Dovesi}},\ }\href
  {https://doi.org/10.1002/jcc.26206} {\bibfield  {journal} {\bibinfo
  {journal} {J. Comput. Chem.}\ }\textbf {\bibinfo {volume} {41}},\ \bibinfo
  {pages} {1638} (\bibinfo {year} {2020})}\BibitemShut {NoStop}%
\bibitem [{\citenamefont {Platonenko}\ \emph {et~al.}(2021)\citenamefont
  {Platonenko}, \citenamefont {Gentile}, \citenamefont {Pascale}, \citenamefont
  {D'Arco},\ and\ \citenamefont {Dovesi}}]{ref:platonenko_C-int}%
  \BibitemOpen
  \bibfield  {author} {\bibinfo {author} {\bibfnamefont {A.}~\bibnamefont
  {Platonenko}}, \bibinfo {author} {\bibfnamefont {F.~S.}\ \bibnamefont
  {Gentile}}, \bibinfo {author} {\bibfnamefont {F.}~\bibnamefont {Pascale}},
  \bibinfo {author} {\bibfnamefont {P.}~\bibnamefont {D'Arco}},\ and\ \bibinfo
  {author} {\bibfnamefont {R.}~\bibnamefont {Dovesi}},\ }\href
  {https://doi.org/10.1002/jcc.26500} {\bibfield  {journal} {\bibinfo
  {journal} {J. Comput. Chem.}\ }\textbf {\bibinfo {volume} {42}},\ \bibinfo
  {pages} {806} (\bibinfo {year} {2021})}\BibitemShut {NoStop}%
\bibitem [{\citenamefont {Simha}\ \emph {et~al.}(2023)\citenamefont {Simha},
  \citenamefont {Herrero-Saboya}, \citenamefont {Giacomazzi}, \citenamefont
  {Martin-Samos}, \citenamefont {Hemeryck},\ and\ \citenamefont
  {Richard}}]{ref:simha2023}%
  \BibitemOpen
  \bibfield  {author} {\bibinfo {author} {\bibfnamefont {C.}~\bibnamefont
  {Simha}}, \bibinfo {author} {\bibfnamefont {G.}~\bibnamefont
  {Herrero-Saboya}}, \bibinfo {author} {\bibfnamefont {L.}~\bibnamefont
  {Giacomazzi}}, \bibinfo {author} {\bibfnamefont {L.}~\bibnamefont
  {Martin-Samos}}, \bibinfo {author} {\bibfnamefont {A.}~\bibnamefont
  {Hemeryck}},\ and\ \bibinfo {author} {\bibfnamefont {N.}~\bibnamefont
  {Richard}},\ }\href {https://doi.org/10.3390/nano13142123} {\bibfield
  {journal} {\bibinfo  {journal} {Nanomaterials}\ }\textbf {\bibinfo {volume}
  {13}},\ \bibinfo {pages} {2123.} (\bibinfo {year} {2023})}\BibitemShut
  {NoStop}%
\bibitem [{\citenamefont {Henkelman}\ \emph {et~al.}(2000)\citenamefont
  {Henkelman}, \citenamefont {Uberuaga},\ and\ \citenamefont
  {J{\'{o}}nsson}}]{ref:Henkelman2000_climbing-neb}%
  \BibitemOpen
  \bibfield  {author} {\bibinfo {author} {\bibfnamefont {G.}~\bibnamefont
  {Henkelman}}, \bibinfo {author} {\bibfnamefont {B.~P.}\ \bibnamefont
  {Uberuaga}},\ and\ \bibinfo {author} {\bibfnamefont {H.}~\bibnamefont
  {J{\'{o}}nsson}},\ }\href {https://doi.org/10.1063/1.1329672} {\bibfield
  {journal} {\bibinfo  {journal} {J. Chem. Phys.}\ }\textbf {\bibinfo {volume}
  {113}},\ \bibinfo {pages} {9901} (\bibinfo {year} {2000})}\BibitemShut
  {NoStop}%
\bibitem [{\citenamefont {Jones}\ and\ \citenamefont
  {Gunnarsson}(1989)}]{ref:constr-occ-delta-scf}%
  \BibitemOpen
  \bibfield  {author} {\bibinfo {author} {\bibfnamefont {R.~O.}\ \bibnamefont
  {Jones}}\ and\ \bibinfo {author} {\bibfnamefont {O.}~\bibnamefont
  {Gunnarsson}},\ }\href {https://doi.org/10.1103/RevModPhys.61.689} {\bibfield
   {journal} {\bibinfo  {journal} {Rev. Mod. Phys.}\ }\textbf {\bibinfo
  {volume} {61}},\ \bibinfo {pages} {689} (\bibinfo {year} {1989})}\BibitemShut
  {NoStop}%
\bibitem [{\citenamefont {Huang}\ and\ \citenamefont
  {Rhys}(1950)}]{ref:huang-rhys-orig}%
  \BibitemOpen
  \bibfield  {author} {\bibinfo {author} {\bibfnamefont {K.}~\bibnamefont
  {Huang}}\ and\ \bibinfo {author} {\bibfnamefont {A.}~\bibnamefont {Rhys}},\
  }\href {https://doi.org/10.1098/rspa.1950.0184} {\bibfield  {journal}
  {\bibinfo  {journal} {Proc. R. Soc. Lond. A}\ }\textbf {\bibinfo {volume}
  {204}},\ \bibinfo {pages} {406} (\bibinfo {year} {1950})}\BibitemShut
  {NoStop}%
\bibitem [{\citenamefont {Alkauskas}\ \emph {et~al.}(2012)\citenamefont
  {Alkauskas}, \citenamefont {Lyons}, \citenamefont {Steiauf},\ and\
  \citenamefont {{Van de Walle}}}]{ref:alkauskas2012_huang-rhys}%
  \BibitemOpen
  \bibfield  {author} {\bibinfo {author} {\bibfnamefont {A.}~\bibnamefont
  {Alkauskas}}, \bibinfo {author} {\bibfnamefont {J.~L.}\ \bibnamefont
  {Lyons}}, \bibinfo {author} {\bibfnamefont {D.}~\bibnamefont {Steiauf}},\
  and\ \bibinfo {author} {\bibfnamefont {C.~G.}\ \bibnamefont {{Van de
  Walle}}},\ }\href {https://doi.org/10.1103/PhysRevLett.109.267401} {\bibfield
   {journal} {\bibinfo  {journal} {Phys. Rev. Lett.}\ }\textbf {\bibinfo
  {volume} {109}},\ \bibinfo {pages} {267401} (\bibinfo {year}
  {2012})}\BibitemShut {NoStop}%
\bibitem [{\citenamefont {Alkauskas}\ \emph {et~al.}(2014)\citenamefont
  {Alkauskas}, \citenamefont {Buckley}, \citenamefont {Awschalom},\ and\
  \citenamefont {{Van de Walle}}}]{ref:alkauskas2014_hr-dw}%
  \BibitemOpen
  \bibfield  {author} {\bibinfo {author} {\bibfnamefont {A.}~\bibnamefont
  {Alkauskas}}, \bibinfo {author} {\bibfnamefont {B.~B.}\ \bibnamefont
  {Buckley}}, \bibinfo {author} {\bibfnamefont {D.~D.}\ \bibnamefont
  {Awschalom}},\ and\ \bibinfo {author} {\bibfnamefont {C.~G.}\ \bibnamefont
  {{Van de Walle}}},\ }\href {https://doi.org/10.1088/1367-2630/16/7/073026}
  {\bibfield  {journal} {\bibinfo  {journal} {New J. Phys.}\ }\textbf {\bibinfo
  {volume} {16}},\ \bibinfo {pages} {073026} (\bibinfo {year}
  {2014})}\BibitemShut {NoStop}%
\bibitem [{\citenamefont {Xiong}\ \emph {et~al.}(2024)\citenamefont {Xiong},
  \citenamefont {Zheng}, \citenamefont {McBride}, \citenamefont {Zhang},
  \citenamefont {Griffin},\ and\ \citenamefont {Hautier}}]{ref:xiong_hautier}%
  \BibitemOpen
  \bibfield  {author} {\bibinfo {author} {\bibfnamefont {Y.}~\bibnamefont
  {Xiong}}, \bibinfo {author} {\bibfnamefont {J.}~\bibnamefont {Zheng}},
  \bibinfo {author} {\bibfnamefont {S.}~\bibnamefont {McBride}}, \bibinfo
  {author} {\bibfnamefont {X.}~\bibnamefont {Zhang}}, \bibinfo {author}
  {\bibfnamefont {S.~M.}\ \bibnamefont {Griffin}},\ and\ \bibinfo {author}
  {\bibfnamefont {G.}~\bibnamefont {Hautier}},\ }\href
  {https://doi.org/10.1021/jacs.4c06613} {\bibfield  {journal} {\bibinfo
  {journal} {J. Am. Chem. Soc.}\ }\textbf {\bibinfo {volume} {146}},\ \bibinfo
  {pages} {30046} (\bibinfo {year} {2024})}\BibitemShut {NoStop}%
\bibitem [{\citenamefont {Turiansky}\ and\ \citenamefont
  {Lyons}(2025)}]{ref:turiansky2025}%
  \BibitemOpen
  \bibfield  {author} {\bibinfo {author} {\bibfnamefont {M.~E.}\ \bibnamefont
  {Turiansky}}\ and\ \bibinfo {author} {\bibfnamefont {J.~L.}\ \bibnamefont
  {Lyons}},\ }\href {https://arxiv.org/abs/2506.12174} {\bibinfo {title}
  {Approximate excited-state potential energy surfaces for defects in solids}}
  (\bibinfo {year} {2025}),\ \Eprint {https://arxiv.org/abs/2506.12174}
  {arXiv:2506.12174 [cond-mat.mtrl-sci]} \BibitemShut {NoStop}%
\bibitem [{Note1()}]{Note1}%
  \BibitemOpen
  \bibinfo {note} {Supercell-size dependence for the T center was also observed
  in Ref.~{\protect \rev@citealp {ref:alaerts2025}}, which calculated exciton
  binding energies and dipole moment changes}\BibitemShut {NoStop}%
\bibitem [{\citenamefont {Karaiskaj}\ \emph {et~al.}(2002)\citenamefont
  {Karaiskaj}, \citenamefont {Thewalt}, \citenamefont {Ruf}, \citenamefont
  {Cardona},\ and\ \citenamefont {Konuma}}]{ref:karaiskaj2002_si-zpr}%
  \BibitemOpen
  \bibfield  {author} {\bibinfo {author} {\bibfnamefont {D.}~\bibnamefont
  {Karaiskaj}}, \bibinfo {author} {\bibfnamefont {M.~L.~W.}\ \bibnamefont
  {Thewalt}}, \bibinfo {author} {\bibfnamefont {T.}~\bibnamefont {Ruf}},
  \bibinfo {author} {\bibfnamefont {M.}~\bibnamefont {Cardona}},\ and\ \bibinfo
  {author} {\bibfnamefont {M.}~\bibnamefont {Konuma}},\ }\href
  {https://doi.org/10.1016/S0038-1098(02)00249-1} {\bibfield  {journal}
  {\bibinfo  {journal} {Solid State Commun.}\ }\textbf {\bibinfo {volume}
  {123}},\ \bibinfo {pages} {87} (\bibinfo {year} {2002})}\BibitemShut
  {NoStop}%
\bibitem [{\citenamefont {Paschotta}(2005)}]{ref:telecom-region}%
  \BibitemOpen
  \bibfield  {author} {\bibinfo {author} {\bibfnamefont {R.}~\bibnamefont
  {Paschotta}},\ }\href {https://doi.org/10.61835/9ec} {\bibinfo {title}
  {Optical fiber communications}},\ \bibinfo {howpublished} {RP Photonics
  Encyclopedia} (\bibinfo {year} {2005})\BibitemShut {NoStop}%
\bibitem [{\citenamefont {Davies}(1989)}]{ref:davies89}%
  \BibitemOpen
  \bibfield  {author} {\bibinfo {author} {\bibfnamefont {G.}~\bibnamefont
  {Davies}},\ }\href {https://doi.org/10.1016/0370-1573(89)90064-1} {\bibfield
  {journal} {\bibinfo  {journal} {Phys. Rep.}\ }\textbf {\bibinfo {volume}
  {176}},\ \bibinfo {pages} {83} (\bibinfo {year} {1989})}\BibitemShut
  {NoStop}%
\bibitem [{\citenamefont {Conzelmann}\ \emph {et~al.}(1983)\citenamefont
  {Conzelmann}, \citenamefont {Graff},\ and\ \citenamefont
  {Weber}}]{ref:conzelmann1983}%
  \BibitemOpen
  \bibfield  {author} {\bibinfo {author} {\bibfnamefont {H.}~\bibnamefont
  {Conzelmann}}, \bibinfo {author} {\bibfnamefont {K.}~\bibnamefont {Graff}},\
  and\ \bibinfo {author} {\bibfnamefont {E.~R.}\ \bibnamefont {Weber}},\ }\href
  {https://doi.org/10.1007/BF00620536} {\bibfield  {journal} {\bibinfo
  {journal} {Applied Physics A Solids and Surfaces}\ }\textbf {\bibinfo
  {volume} {30}},\ \bibinfo {pages} {169} (\bibinfo {year} {1983})}\BibitemShut
  {NoStop}%
\bibitem [{\citenamefont {D{\"{o}}rnen}\ \emph {et~al.}(1985)\citenamefont
  {D{\"{o}}rnen}, \citenamefont {Sauer},\ and\ \citenamefont
  {Pensl}}]{ref:dornen85}%
  \BibitemOpen
  \bibfield  {author} {\bibinfo {author} {\bibfnamefont {A.}~\bibnamefont
  {D{\"{o}}rnen}}, \bibinfo {author} {\bibfnamefont {R.}~\bibnamefont
  {Sauer}},\ and\ \bibinfo {author} {\bibfnamefont {G.}~\bibnamefont {Pensl}},\
  }\href {https://doi.org/10.1557/PROC-59-545} {\bibfield  {journal} {\bibinfo
  {journal} {MRS Online Proc. Libr.}\ }\textbf {\bibinfo {volume} {59}},\
  \bibinfo {pages} {545} (\bibinfo {year} {1985})}\BibitemShut {NoStop}%
\bibitem [{\citenamefont {D{\"{o}}rnen}\ \emph
  {et~al.}(1986{\natexlab{a}})\citenamefont {D{\"{o}}rnen}, \citenamefont
  {Pensl},\ and\ \citenamefont {Sauer}}]{ref:dornen86-prb}%
  \BibitemOpen
  \bibfield  {author} {\bibinfo {author} {\bibfnamefont {A.}~\bibnamefont
  {D{\"{o}}rnen}}, \bibinfo {author} {\bibfnamefont {G.}~\bibnamefont
  {Pensl}},\ and\ \bibinfo {author} {\bibfnamefont {R.}~\bibnamefont {Sauer}},\
  }\href {https://doi.org/10.1103/PhysRevB.33.1495} {\bibfield  {journal}
  {\bibinfo  {journal} {Phys. Rev. B}\ }\textbf {\bibinfo {volume} {33}},\
  \bibinfo {pages} {1495(R)} (\bibinfo {year}
  {1986}{\natexlab{a}})}\BibitemShut {NoStop}%
\bibitem [{\citenamefont {D{\"{o}}rnen}\ \emph
  {et~al.}(1986{\natexlab{b}})\citenamefont {D{\"{o}}rnen}, \citenamefont
  {Pensl},\ and\ \citenamefont {Sauer}}]{ref:dornen86}%
  \BibitemOpen
  \bibfield  {author} {\bibinfo {author} {\bibfnamefont {A.}~\bibnamefont
  {D{\"{o}}rnen}}, \bibinfo {author} {\bibfnamefont {G.}~\bibnamefont
  {Pensl}},\ and\ \bibinfo {author} {\bibfnamefont {R.}~\bibnamefont {Sauer}},\
  }\href {https://doi.org/10.1016/0038-1098(86)90167-5} {\bibfield  {journal}
  {\bibinfo  {journal} {Solid State Commun.}\ }\textbf {\bibinfo {volume}
  {57}},\ \bibinfo {pages} {861} (\bibinfo {year}
  {1986}{\natexlab{b}})}\BibitemShut {NoStop}%
\bibitem [{\citenamefont {D{\"{o}}rnen}\ \emph {et~al.}(1987)\citenamefont
  {D{\"{o}}rnen}, \citenamefont {Pensl},\ and\ \citenamefont
  {Sauer}}]{ref:dornen87}%
  \BibitemOpen
  \bibfield  {author} {\bibinfo {author} {\bibfnamefont {A.}~\bibnamefont
  {D{\"{o}}rnen}}, \bibinfo {author} {\bibfnamefont {G.}~\bibnamefont
  {Pensl}},\ and\ \bibinfo {author} {\bibfnamefont {R.}~\bibnamefont {Sauer}},\
  }\href {https://doi.org/10.1103/PhysRevB.35.9318} {\bibfield  {journal}
  {\bibinfo  {journal} {Phys. Rev. B}\ }\textbf {\bibinfo {volume} {35}},\
  \bibinfo {pages} {9318} (\bibinfo {year} {1987})}\BibitemShut {NoStop}%
\bibitem [{\citenamefont {D{\"{o}}rnen}\ \emph {et~al.}(1988)\citenamefont
  {D{\"{o}}rnen}, \citenamefont {Sauer},\ and\ \citenamefont
  {Pensl}}]{ref:dornen88}%
  \BibitemOpen
  \bibfield  {author} {\bibinfo {author} {\bibfnamefont {A.}~\bibnamefont
  {D{\"{o}}rnen}}, \bibinfo {author} {\bibfnamefont {R.}~\bibnamefont
  {Sauer}},\ and\ \bibinfo {author} {\bibfnamefont {G.}~\bibnamefont {Pensl}},\
  }\href {https://doi.org/10.1007/BF02652141} {\bibfield  {journal} {\bibinfo
  {journal} {J. Electron. Mater.}\ }\textbf {\bibinfo {volume} {17}},\ \bibinfo
  {pages} {121} (\bibinfo {year} {1988})}\BibitemShut {NoStop}%
\bibitem [{\citenamefont {Stoneham}(1975)}]{ref:stoneham}%
  \BibitemOpen
  \bibfield  {author} {\bibinfo {author} {\bibfnamefont {A.~M.}\ \bibnamefont
  {Stoneham}},\ }\href@noop {} {\emph {\bibinfo {title} {{Theory of Defects in
  Solids: Electronic Structure of Defects in Insulators and Semiconductors,
  Monographs on the Physics and Chemistry of Materials}}}}\ (\bibinfo
  {publisher} {Clarendon},\ \bibinfo {address} {Oxford},\ \bibinfo {year}
  {1975})\BibitemShut {NoStop}%
\bibitem [{\citenamefont {Weisskopf}\ and\ \citenamefont
  {Wigner}(1930)}]{ref:wig-weiss}%
  \BibitemOpen
  \bibfield  {author} {\bibinfo {author} {\bibfnamefont {V.}~\bibnamefont
  {Weisskopf}}\ and\ \bibinfo {author} {\bibfnamefont {E.}~\bibnamefont
  {Wigner}},\ }\href {https://doi.org/10.1007/BF01336768} {\bibfield  {journal}
  {\bibinfo  {journal} {Zeitschrift f{\"{u}}r Phys.}\ }\textbf {\bibinfo
  {volume} {63}},\ \bibinfo {pages} {54} (\bibinfo {year} {1930})}\BibitemShut
  {NoStop}%
\bibitem [{\citenamefont {Razinkovas}\ \emph {et~al.}(2021)\citenamefont
  {Razinkovas}, \citenamefont {Maciaszek}, \citenamefont {Reinhard},
  \citenamefont {Doherty},\ and\ \citenamefont {Alkauskas}}]{ref:razinkovas}%
  \BibitemOpen
  \bibfield  {author} {\bibinfo {author} {\bibfnamefont {L.}~\bibnamefont
  {Razinkovas}}, \bibinfo {author} {\bibfnamefont {M.}~\bibnamefont
  {Maciaszek}}, \bibinfo {author} {\bibfnamefont {F.}~\bibnamefont {Reinhard}},
  \bibinfo {author} {\bibfnamefont {M.~W.}\ \bibnamefont {Doherty}},\ and\
  \bibinfo {author} {\bibfnamefont {A.}~\bibnamefont {Alkauskas}},\ }\href
  {https://doi.org/10.1103/PhysRevB.104.235301} {\bibfield  {journal} {\bibinfo
   {journal} {Phys. Rev. B}\ }\textbf {\bibinfo {volume} {104}},\ \bibinfo
  {pages} {235301} (\bibinfo {year} {2021})}\BibitemShut {NoStop}%
\bibitem [{Note2()}]{Note2}%
  \BibitemOpen
  \bibinfo {note} {The scaling method here results in a radiative lifetime
  formula that is is analogous to the radiative capture rate formalism~{\cite
  {ref:dreyer}} as explained in Sec.~{\ref {sec:supp_capture}} in the SM~{\cite
  {ref:supplemental}}}\BibitemShut {NoStop}%
\bibitem [{\citenamefont {Kazemi}\ \emph {et~al.}(2025)\citenamefont {Kazemi},
  \citenamefont {Keshavarz}, \citenamefont {Turiansky}, \citenamefont {Lyons},
  \citenamefont {Abrosimov}, \citenamefont {Simmons}, \citenamefont
  {Higginbottom},\ and\ \citenamefont {Thewalt}}]{ref:kazemi2025}%
  \BibitemOpen
  \bibfield  {author} {\bibinfo {author} {\bibfnamefont {M.}~\bibnamefont
  {Kazemi}}, \bibinfo {author} {\bibfnamefont {M.}~\bibnamefont {Keshavarz}},
  \bibinfo {author} {\bibfnamefont {M.~E.}\ \bibnamefont {Turiansky}}, \bibinfo
  {author} {\bibfnamefont {J.~L.}\ \bibnamefont {Lyons}}, \bibinfo {author}
  {\bibfnamefont {N.~V.}\ \bibnamefont {Abrosimov}}, \bibinfo {author}
  {\bibfnamefont {S.}~\bibnamefont {Simmons}}, \bibinfo {author} {\bibfnamefont
  {D.~B.}\ \bibnamefont {Higginbottom}},\ and\ \bibinfo {author} {\bibfnamefont
  {M.~L.~W.}\ \bibnamefont {Thewalt}},\ }\href
  {https://arxiv.org/abs/2510.23862} {\bibinfo {title} {Giant isotope effect on
  the excited-state lifetime and emission efficiency of the silicon t centre}}
  (\bibinfo {year} {2025}),\ \Eprint {https://arxiv.org/abs/2510.23862}
  {arXiv:2510.23862 [quant-ph]} \BibitemShut {NoStop}%
\bibitem [{\citenamefont {Boerner}\ \emph {et~al.}(2023)\citenamefont
  {Boerner}, \citenamefont {Deems}, \citenamefont {Furlani}, \citenamefont
  {Knuth},\ and\ \citenamefont {Towns}}]{ref:access}%
  \BibitemOpen
  \bibfield  {author} {\bibinfo {author} {\bibfnamefont {T.~J.}\ \bibnamefont
  {Boerner}}, \bibinfo {author} {\bibfnamefont {S.}~\bibnamefont {Deems}},
  \bibinfo {author} {\bibfnamefont {T.~R.}\ \bibnamefont {Furlani}}, \bibinfo
  {author} {\bibfnamefont {S.~L.}\ \bibnamefont {Knuth}},\ and\ \bibinfo
  {author} {\bibfnamefont {J.}~\bibnamefont {Towns}},\ }in\ \href
  {https://doi.org/10.1145/3569951.3597559} {\emph {\bibinfo {booktitle}
  {Pract. Exp. Adv. Res. Comput.}}},\ \bibinfo {series and number} {PEARC '23}\
  (\bibinfo  {publisher} {Association for Computing Machinery},\ \bibinfo
  {address} {New York, NY, USA},\ \bibinfo {year} {2023})\ pp.\ \bibinfo
  {pages} {173--176}\BibitemShut {NoStop}%
\bibitem [{\citenamefont {Schultz}\ and\ \citenamefont
  {Nelson}(2001)}]{ref:schultz2001}%
  \BibitemOpen
  \bibfield  {author} {\bibinfo {author} {\bibfnamefont {P.~A.}\ \bibnamefont
  {Schultz}}\ and\ \bibinfo {author} {\bibfnamefont {J.~S.}\ \bibnamefont
  {Nelson}},\ }\href {https://doi.org/10.1063/1.1345828} {\bibfield  {journal}
  {\bibinfo  {journal} {Appl. Phys. Lett.}\ }\textbf {\bibinfo {volume} {78}},\
  \bibinfo {pages} {736} (\bibinfo {year} {2001})}\BibitemShut {NoStop}%
\bibitem [{\citenamefont {Wannier}(1937)}]{ref:wannier}%
  \BibitemOpen
  \bibfield  {author} {\bibinfo {author} {\bibfnamefont {G.~H.}\ \bibnamefont
  {Wannier}},\ }\href {https://doi.org/10.1103/PhysRev.52.191} {\bibfield
  {journal} {\bibinfo  {journal} {Phys. Rev.}\ }\textbf {\bibinfo {volume}
  {52}},\ \bibinfo {pages} {191} (\bibinfo {year} {1937})}\BibitemShut
  {NoStop}%
\bibitem [{\citenamefont {Luttinger}\ and\ \citenamefont
  {Kohn}(1955)}]{ref:kohn-luttinger}%
  \BibitemOpen
  \bibfield  {author} {\bibinfo {author} {\bibfnamefont {J.~M.}\ \bibnamefont
  {Luttinger}}\ and\ \bibinfo {author} {\bibfnamefont {W.}~\bibnamefont
  {Kohn}},\ }\href {https://doi.org/10.1103/PhysRev.97.869} {\bibfield
  {journal} {\bibinfo  {journal} {Phys. Rev.}\ }\textbf {\bibinfo {volume}
  {97}},\ \bibinfo {pages} {869} (\bibinfo {year} {1955})}\BibitemShut
  {NoStop}%
\bibitem [{\citenamefont {P{\"{a}}ssler}(1976)}]{ref:passler}%
  \BibitemOpen
  \bibfield  {author} {\bibinfo {author} {\bibfnamefont {R.}~\bibnamefont
  {P{\"{a}}ssler}},\ }\href {https://doi.org/10.1002/pssb.2220780222}
  {\bibfield  {journal} {\bibinfo  {journal} {Phys. Status Solidi}\ }\textbf
  {\bibinfo {volume} {78}},\ \bibinfo {pages} {625} (\bibinfo {year}
  {1976})}\BibitemShut {NoStop}%
\bibitem [{\citenamefont
  {Griffiths}(2005{\natexlab{a}})}]{ref:griffiths_rydberg}%
  \BibitemOpen
  \bibfield  {author} {\bibinfo {author} {\bibfnamefont {D.~J.}\ \bibnamefont
  {Griffiths}},\ }\href@noop {} {\emph {\bibinfo {title} {{Introduction to
  Quantum Mechanics}}}},\ \bibinfo {edition} {2nd}\ ed.\ (\bibinfo  {publisher}
  {Pearson Education},\ \bibinfo {year} {2005})\ p.\ \bibinfo {pages}
  {158}\BibitemShut {NoStop}%
\bibitem [{\citenamefont {Dunlap}\ and\ \citenamefont
  {Watters}(1953)}]{ref:Si_dielConst}%
  \BibitemOpen
  \bibfield  {author} {\bibinfo {author} {\bibfnamefont {W.~C.}\ \bibnamefont
  {Dunlap}}\ and\ \bibinfo {author} {\bibfnamefont {R.~L.}\ \bibnamefont
  {Watters}},\ }\href {https://doi.org/10.1103/PhysRev.92.1396} {\bibfield
  {journal} {\bibinfo  {journal} {Phys. Rev.}\ }\textbf {\bibinfo {volume}
  {92}},\ \bibinfo {pages} {1396} (\bibinfo {year} {1953})}\BibitemShut
  {NoStop}%
\bibitem [{\citenamefont {Conzelmann}(1987)}]{ref:conzelmann1987}%
  \BibitemOpen
  \bibfield  {author} {\bibinfo {author} {\bibfnamefont {H.}~\bibnamefont
  {Conzelmann}},\ }\href {https://doi.org/10.1007/BF00618154} {\bibfield
  {journal} {\bibinfo  {journal} {Appl. Phys. A}\ }\textbf {\bibinfo {volume}
  {42}},\ \bibinfo {pages} {1} (\bibinfo {year} {1987})}\BibitemShut {NoStop}%
\bibitem [{\citenamefont {Sommerfeld}(1931)}]{ref:sommerfeld}%
  \BibitemOpen
  \bibfield  {author} {\bibinfo {author} {\bibfnamefont {A.}~\bibnamefont
  {Sommerfeld}},\ }\href
  {https://doi.org/https://doi.org/10.1002/andp.19314030302} {\bibfield
  {journal} {\bibinfo  {journal} {Ann. Phys.}\ }\textbf {\bibinfo {volume}
  {403}},\ \bibinfo {pages} {257} (\bibinfo {year} {1931})}\BibitemShut
  {NoStop}%
\bibitem [{\citenamefont {Turiansky}\ \emph
  {et~al.}(2024{\natexlab{b}})\citenamefont {Turiansky}, \citenamefont
  {Alkauskas},\ and\ \citenamefont {{Van de Walle}}}]{ref:turiansky_jpcm2024}%
  \BibitemOpen
  \bibfield  {author} {\bibinfo {author} {\bibfnamefont {M.~E.}\ \bibnamefont
  {Turiansky}}, \bibinfo {author} {\bibfnamefont {A.}~\bibnamefont
  {Alkauskas}},\ and\ \bibinfo {author} {\bibfnamefont {C.~G.}\ \bibnamefont
  {{Van de Walle}}},\ }\href {https://doi.org/10.1088/1361-648X/ad2588}
  {\bibfield  {journal} {\bibinfo  {journal} {J. Phys. Condens. Matter}\
  }\textbf {\bibinfo {volume} {36}},\ \bibinfo {pages} {195902} (\bibinfo
  {year} {2024}{\natexlab{b}})}\BibitemShut {NoStop}%
\bibitem [{\citenamefont
  {Griffiths}(2005{\natexlab{b}})}]{ref:griffiths_hyd-1s-wf}%
  \BibitemOpen
  \bibfield  {author} {\bibinfo {author} {\bibfnamefont {D.~J.}\ \bibnamefont
  {Griffiths}},\ }\href@noop {} {\emph {\bibinfo {title} {{Introduction to
  Quantum Mechanics}}}},\ \bibinfo {edition} {2nd}\ ed.\ (\bibinfo  {publisher}
  {Pearson Education},\ \bibinfo {year} {2005})\ p.\ \bibinfo {pages}
  {151}\BibitemShut {NoStop}%
\bibitem [{\citenamefont {Alaerts}\ \emph {et~al.}(2025)\citenamefont
  {Alaerts}, \citenamefont {Xiong}, \citenamefont {Griffin},\ and\
  \citenamefont {Hautier}}]{ref:alaerts2025}%
  \BibitemOpen
  \bibfield  {author} {\bibinfo {author} {\bibfnamefont {L.}~\bibnamefont
  {Alaerts}}, \bibinfo {author} {\bibfnamefont {Y.}~\bibnamefont {Xiong}},
  \bibinfo {author} {\bibfnamefont {S.~M.}\ \bibnamefont {Griffin}},\ and\
  \bibinfo {author} {\bibfnamefont {G.}~\bibnamefont {Hautier}},\ }\href
  {https://doi.org/10.1103/b59h-4wcn} {\bibfield  {journal} {\bibinfo
  {journal} {Phys. Rev. B}\ }\textbf {\bibinfo {volume} {112}},\ \bibinfo
  {pages} {125114} (\bibinfo {year} {2025})}\BibitemShut {NoStop}%
\bibitem [{\citenamefont {Dreyer}\ \emph {et~al.}(2020)\citenamefont {Dreyer},
  \citenamefont {Alkauskas}, \citenamefont {Lyons},\ and\ \citenamefont {{Van
  de Walle}}}]{ref:dreyer}%
  \BibitemOpen
  \bibfield  {author} {\bibinfo {author} {\bibfnamefont {C.~E.}\ \bibnamefont
  {Dreyer}}, \bibinfo {author} {\bibfnamefont {A.}~\bibnamefont {Alkauskas}},
  \bibinfo {author} {\bibfnamefont {J.~L.}\ \bibnamefont {Lyons}},\ and\
  \bibinfo {author} {\bibfnamefont {C.~G.}\ \bibnamefont {{Van de Walle}}},\
  }\href {https://doi.org/10.1103/PhysRevB.102.085305} {\bibfield  {journal}
  {\bibinfo  {journal} {Phys. Rev. B}\ }\textbf {\bibinfo {volume} {102}},\
  \bibinfo {pages} {085305} (\bibinfo {year} {2020})}\BibitemShut {NoStop}%
\end{thebibliography}%


\begin{thebibliography}{27}%
\makeatletter
\providecommand \@ifxundefined [1]{%
 \@ifx{#1\undefined}
}%
\providecommand \@ifnum [1]{%
 \ifnum #1\expandafter \@firstoftwo
 \else \expandafter \@secondoftwo
 \fi
}%
\providecommand \@ifx [1]{%
 \ifx #1\expandafter \@firstoftwo
 \else \expandafter \@secondoftwo
 \fi
}%
\providecommand \natexlab [1]{#1}%
\providecommand \enquote  [1]{``#1''}%
\providecommand \bibnamefont  [1]{#1}%
\providecommand \bibfnamefont [1]{#1}%
\providecommand \citenamefont [1]{#1}%
\providecommand \href@noop [0]{\@secondoftwo}%
\providecommand \href [0]{\begingroup \@sanitize@url \@href}%
\providecommand \@href[1]{\@@startlink{#1}\@@href}%
\providecommand \@@href[1]{\endgroup#1\@@endlink}%
\providecommand \@sanitize@url [0]{\catcode `\\12\catcode `\$12\catcode
  `\&12\catcode `\#12\catcode `\^12\catcode `\_12\catcode `\%12\relax}%
\providecommand \@@startlink[1]{}%
\providecommand \@@endlink[0]{}%
\providecommand \url  [0]{\begingroup\@sanitize@url \@url }%
\providecommand \@url [1]{\endgroup\@href {#1}{\urlprefix }}%
\providecommand \urlprefix  [0]{URL }%
\providecommand \Eprint [0]{\href }%
\providecommand \doibase [0]{http://dx.doi.org/}%
\providecommand \selectlanguage [0]{\@gobble}%
\providecommand \bibinfo  [0]{\@secondoftwo}%
\providecommand \bibfield  [0]{\@secondoftwo}%
\providecommand \translation [1]{[#1]}%
\providecommand \BibitemOpen [0]{}%
\providecommand \bibitemStop [0]{}%
\providecommand \bibitemNoStop [0]{.\EOS\space}%
\providecommand \EOS [0]{\spacefactor3000\relax}%
\providecommand \BibitemShut  [1]{\csname bibitem#1\endcsname}%
\let\auto@bib@innerbib\@empty
\bibitem [{\citenamefont {Tersoff}(1990)}]{ref:tersoff1990}%
  \BibitemOpen
  \bibfield  {author} {\bibinfo {author} {\bibfnamefont {J.}~\bibnamefont
  {Tersoff}},\ }\href {\doibase 10.1103/PhysRevLett.64.1757} {\bibfield
  {journal} {\bibinfo  {journal} {Phys. Rev. Lett.}\ }\textbf {\bibinfo
  {volume} {64}},\ \bibinfo {pages} {1757} (\bibinfo {year}
  {1990})}\BibitemShut {NoStop}%
\bibitem [{\citenamefont {Platonenko}\ \emph {et~al.}(2019)\citenamefont
  {Platonenko}, \citenamefont {Gentile}, \citenamefont {Maul}, \citenamefont
  {Pascale}, \citenamefont {Kotomin},\ and\ \citenamefont
  {Dovesi}}]{ref:platonenko_N-int}%
  \BibitemOpen
  \bibfield  {author} {\bibinfo {author} {\bibfnamefont {A.}~\bibnamefont
  {Platonenko}}, \bibinfo {author} {\bibfnamefont {F.~S.}\ \bibnamefont
  {Gentile}}, \bibinfo {author} {\bibfnamefont {J.}~\bibnamefont {Maul}},
  \bibinfo {author} {\bibfnamefont {F.}~\bibnamefont {Pascale}}, \bibinfo
  {author} {\bibfnamefont {E.~A.}\ \bibnamefont {Kotomin}}, \ and\ \bibinfo
  {author} {\bibfnamefont {R.}~\bibnamefont {Dovesi}},\ }\href {\doibase
  10.1016/j.mtcomm.2019.100616} {\bibfield  {journal} {\bibinfo  {journal}
  {Mater. Today Commun.}\ }\textbf {\bibinfo {volume} {21}},\ \bibinfo {pages}
  {100616} (\bibinfo {year} {2019})}\BibitemShut {NoStop}%
\bibitem [{\citenamefont {Gentile}\ \emph {et~al.}(2020)\citenamefont
  {Gentile}, \citenamefont {Platonenko}, \citenamefont {El-Kelany},
  \citenamefont {R{\'{e}}rat}, \citenamefont {D'Arco},\ and\ \citenamefont
  {Dovesi}}]{ref:platonenko_C-sub}%
  \BibitemOpen
  \bibfield  {author} {\bibinfo {author} {\bibfnamefont {F.~S.}\ \bibnamefont
  {Gentile}}, \bibinfo {author} {\bibfnamefont {A.}~\bibnamefont {Platonenko}},
  \bibinfo {author} {\bibfnamefont {K.~E.}\ \bibnamefont {El-Kelany}}, \bibinfo
  {author} {\bibfnamefont {M.}~\bibnamefont {R{\'{e}}rat}}, \bibinfo {author}
  {\bibfnamefont {P.}~\bibnamefont {D'Arco}}, \ and\ \bibinfo {author}
  {\bibfnamefont {R.}~\bibnamefont {Dovesi}},\ }\href {\doibase
  10.1002/jcc.26206} {\bibfield  {journal} {\bibinfo  {journal} {J. Comput.
  Chem.}\ }\textbf {\bibinfo {volume} {41}},\ \bibinfo {pages} {1638} (\bibinfo
  {year} {2020})}\BibitemShut {NoStop}%
\bibitem [{\citenamefont {Platonenko}\ \emph {et~al.}(2021)\citenamefont
  {Platonenko}, \citenamefont {Gentile}, \citenamefont {Pascale}, \citenamefont
  {D'Arco},\ and\ \citenamefont {Dovesi}}]{ref:platonenko_C-int}%
  \BibitemOpen
  \bibfield  {author} {\bibinfo {author} {\bibfnamefont {A.}~\bibnamefont
  {Platonenko}}, \bibinfo {author} {\bibfnamefont {F.~S.}\ \bibnamefont
  {Gentile}}, \bibinfo {author} {\bibfnamefont {F.}~\bibnamefont {Pascale}},
  \bibinfo {author} {\bibfnamefont {P.}~\bibnamefont {D'Arco}}, \ and\ \bibinfo
  {author} {\bibfnamefont {R.}~\bibnamefont {Dovesi}},\ }\href {\doibase
  10.1002/jcc.26500} {\bibfield  {journal} {\bibinfo  {journal} {J. Comput.
  Chem.}\ }\textbf {\bibinfo {volume} {42}},\ \bibinfo {pages} {806} (\bibinfo
  {year} {2021})}\BibitemShut {NoStop}%
\bibitem [{\citenamefont {Simha}\ \emph {et~al.}(2023)\citenamefont {Simha},
  \citenamefont {Herrero-Saboya}, \citenamefont {Giacomazzi}, \citenamefont
  {Martin-Samos}, \citenamefont {Hemeryck},\ and\ \citenamefont
  {Richard}}]{ref:simha2023}%
  \BibitemOpen
  \bibfield  {author} {\bibinfo {author} {\bibfnamefont {C.}~\bibnamefont
  {Simha}}, \bibinfo {author} {\bibfnamefont {G.}~\bibnamefont
  {Herrero-Saboya}}, \bibinfo {author} {\bibfnamefont {L.}~\bibnamefont
  {Giacomazzi}}, \bibinfo {author} {\bibfnamefont {L.}~\bibnamefont
  {Martin-Samos}}, \bibinfo {author} {\bibfnamefont {A.}~\bibnamefont
  {Hemeryck}}, \ and\ \bibinfo {author} {\bibfnamefont {N.}~\bibnamefont
  {Richard}},\ }\href {\doibase 10.3390/nano13142123} {\bibfield  {journal}
  {\bibinfo  {journal} {Nanomaterials}\ }\textbf {\bibinfo {volume} {13}},\
  \bibinfo {pages} {2123.} (\bibinfo {year} {2023})}\BibitemShut {NoStop}%
\bibitem [{\citenamefont {Schultz}\ and\ \citenamefont
  {Nelson}(2001)}]{ref:schultz2001}%
  \BibitemOpen
  \bibfield  {author} {\bibinfo {author} {\bibfnamefont {P.~A.}\ \bibnamefont
  {Schultz}}\ and\ \bibinfo {author} {\bibfnamefont {J.~S.}\ \bibnamefont
  {Nelson}},\ }\href {\doibase 10.1063/1.1345828} {\bibfield  {journal}
  {\bibinfo  {journal} {Appl. Phys. Lett.}\ }\textbf {\bibinfo {volume} {78}},\
  \bibinfo {pages} {736} (\bibinfo {year} {2001})}\BibitemShut {NoStop}%
\bibitem [{\citenamefont {Henkelman}\ \emph {et~al.}(2000)\citenamefont
  {Henkelman}, \citenamefont {Uberuaga},\ and\ \citenamefont
  {J{\'{o}}nsson}}]{ref:Henkelman2000_climbing-neb}%
  \BibitemOpen
  \bibfield  {author} {\bibinfo {author} {\bibfnamefont {G.}~\bibnamefont
  {Henkelman}}, \bibinfo {author} {\bibfnamefont {B.~P.}\ \bibnamefont
  {Uberuaga}}, \ and\ \bibinfo {author} {\bibfnamefont {H.}~\bibnamefont
  {J{\'{o}}nsson}},\ }\href {\doibase 10.1063/1.1329672} {\bibfield  {journal}
  {\bibinfo  {journal} {J. Chem. Phys.}\ }\textbf {\bibinfo {volume} {113}},\
  \bibinfo {pages} {9901} (\bibinfo {year} {2000})}\BibitemShut {NoStop}%
\bibitem [{\citenamefont {Wannier}(1937)}]{ref:wannier}%
  \BibitemOpen
  \bibfield  {author} {\bibinfo {author} {\bibfnamefont {G.~H.}\ \bibnamefont
  {Wannier}},\ }\href {\doibase 10.1103/PhysRev.52.191} {\bibfield  {journal}
  {\bibinfo  {journal} {Phys. Rev.}\ }\textbf {\bibinfo {volume} {52}},\
  \bibinfo {pages} {191} (\bibinfo {year} {1937})}\BibitemShut {NoStop}%
\bibitem [{\citenamefont {Luttinger}\ and\ \citenamefont
  {Kohn}(1955)}]{ref:kohn-luttinger}%
  \BibitemOpen
  \bibfield  {author} {\bibinfo {author} {\bibfnamefont {J.~M.}\ \bibnamefont
  {Luttinger}}\ and\ \bibinfo {author} {\bibfnamefont {W.}~\bibnamefont
  {Kohn}},\ }\href {\doibase 10.1103/PhysRev.97.869} {\bibfield  {journal}
  {\bibinfo  {journal} {Phys. Rev.}\ }\textbf {\bibinfo {volume} {97}},\
  \bibinfo {pages} {869} (\bibinfo {year} {1955})}\BibitemShut {NoStop}%
\bibitem [{\citenamefont {P{\"{a}}ssler}(1976)}]{ref:passler}%
  \BibitemOpen
  \bibfield  {author} {\bibinfo {author} {\bibfnamefont {R.}~\bibnamefont
  {P{\"{a}}ssler}},\ }\href {\doibase 10.1002/pssb.2220780222} {\bibfield
  {journal} {\bibinfo  {journal} {Phys. Status Solidi}\ }\textbf {\bibinfo
  {volume} {78}},\ \bibinfo {pages} {625} (\bibinfo {year} {1976})}\BibitemShut
  {NoStop}%
\bibitem [{\citenamefont
  {Griffiths}(2005{\natexlab{a}})}]{ref:griffiths_rydberg}%
  \BibitemOpen
  \bibfield  {author} {\bibinfo {author} {\bibfnamefont {D.~J.}\ \bibnamefont
  {Griffiths}},\ }\href@noop {} {\emph {\bibinfo {title} {{Introduction to
  Quantum Mechanics}}}},\ \bibinfo {edition} {2nd}\ ed.\ (\bibinfo  {publisher}
  {Pearson Education},\ \bibinfo {year} {2005})\ p.\ \bibinfo {pages}
  {158}\BibitemShut {NoStop}%
\bibitem [{\citenamefont {Shur}(1996)}]{ref:si_exp-lattConst}%
  \BibitemOpen
  \bibfield  {author} {\bibinfo {author} {\bibfnamefont {M.~S.}\ \bibnamefont
  {Shur}},\ }\href@noop {} {\emph {\bibinfo {title} {{Handbook Series on
  Semiconductor Parameters}}}},\ Vol.~\bibinfo {volume} {1}\ (\bibinfo
  {publisher} {World Scientific},\ \bibinfo {year} {1996})\BibitemShut
  {NoStop}%
\bibitem [{\citenamefont {Dunlap}\ and\ \citenamefont
  {Watters}(1953)}]{ref:Si_dielConst}%
  \BibitemOpen
  \bibfield  {author} {\bibinfo {author} {\bibfnamefont {W.~C.}\ \bibnamefont
  {Dunlap}}\ and\ \bibinfo {author} {\bibfnamefont {R.~L.}\ \bibnamefont
  {Watters}},\ }\href {\doibase 10.1103/PhysRev.92.1396} {\bibfield  {journal}
  {\bibinfo  {journal} {Phys. Rev.}\ }\textbf {\bibinfo {volume} {92}},\
  \bibinfo {pages} {1396} (\bibinfo {year} {1953})}\BibitemShut {NoStop}%
\bibitem [{\citenamefont {Bergeron}\ \emph {et~al.}(2020)\citenamefont
  {Bergeron}, \citenamefont {Chartrand}, \citenamefont {Kurkjian},
  \citenamefont {Morse}, \citenamefont {Riemann}, \citenamefont {Abrosimov},
  \citenamefont {Becker}, \citenamefont {Pohl}, \citenamefont {Thewalt},\ and\
  \citenamefont {Simmons}}]{ref:bergeron}%
  \BibitemOpen
  \bibfield  {author} {\bibinfo {author} {\bibfnamefont {L.}~\bibnamefont
  {Bergeron}}, \bibinfo {author} {\bibfnamefont {C.}~\bibnamefont {Chartrand}},
  \bibinfo {author} {\bibfnamefont {A.~T.~K.}\ \bibnamefont {Kurkjian}},
  \bibinfo {author} {\bibfnamefont {K.~J.}\ \bibnamefont {Morse}}, \bibinfo
  {author} {\bibfnamefont {H.}~\bibnamefont {Riemann}}, \bibinfo {author}
  {\bibfnamefont {N.~V.}\ \bibnamefont {Abrosimov}}, \bibinfo {author}
  {\bibfnamefont {P.}~\bibnamefont {Becker}}, \bibinfo {author} {\bibfnamefont
  {H.-J.}\ \bibnamefont {Pohl}}, \bibinfo {author} {\bibfnamefont {M.~L.~W.}\
  \bibnamefont {Thewalt}}, \ and\ \bibinfo {author} {\bibfnamefont
  {S.}~\bibnamefont {Simmons}},\ }\href {\doibase 10.1103/PRXQuantum.1.020301}
  {\bibfield  {journal} {\bibinfo  {journal} {PRX Quantum}\ }\textbf {\bibinfo
  {volume} {1}},\ \bibinfo {pages} {020301} (\bibinfo {year}
  {2020})}\BibitemShut {NoStop}%
\bibitem [{\citenamefont {Alkauskas}\ \emph {et~al.}(2012)\citenamefont
  {Alkauskas}, \citenamefont {Lyons}, \citenamefont {Steiauf},\ and\
  \citenamefont {{Van de Walle}}}]{ref:alkauskas2012_huang-rhys}%
  \BibitemOpen
  \bibfield  {author} {\bibinfo {author} {\bibfnamefont {A.}~\bibnamefont
  {Alkauskas}}, \bibinfo {author} {\bibfnamefont {J.~L.}\ \bibnamefont
  {Lyons}}, \bibinfo {author} {\bibfnamefont {D.}~\bibnamefont {Steiauf}}, \
  and\ \bibinfo {author} {\bibfnamefont {C.~G.}\ \bibnamefont {{Van de
  Walle}}},\ }\href {\doibase 10.1103/PhysRevLett.109.267401} {\bibfield
  {journal} {\bibinfo  {journal} {Phys. Rev. Lett.}\ }\textbf {\bibinfo
  {volume} {109}},\ \bibinfo {pages} {267401} (\bibinfo {year}
  {2012})}\BibitemShut {NoStop}%
\bibitem [{\citenamefont {D{\"{o}}rnen}\ \emph {et~al.}(1985)\citenamefont
  {D{\"{o}}rnen}, \citenamefont {Sauer},\ and\ \citenamefont
  {Pensl}}]{ref:dornen85}%
  \BibitemOpen
  \bibfield  {author} {\bibinfo {author} {\bibfnamefont {A.}~\bibnamefont
  {D{\"{o}}rnen}}, \bibinfo {author} {\bibfnamefont {R.}~\bibnamefont {Sauer}},
  \ and\ \bibinfo {author} {\bibfnamefont {G.}~\bibnamefont {Pensl}},\ }\href
  {\doibase 10.1557/PROC-59-545} {\bibfield  {journal} {\bibinfo  {journal}
  {MRS Online Proc. Libr.}\ }\textbf {\bibinfo {volume} {59}},\ \bibinfo
  {pages} {545} (\bibinfo {year} {1985})}\BibitemShut {NoStop}%
\bibitem [{\citenamefont {D{\"{o}}rnen}\ \emph
  {et~al.}(1986{\natexlab{a}})\citenamefont {D{\"{o}}rnen}, \citenamefont
  {Pensl},\ and\ \citenamefont {Sauer}}]{ref:dornen86-prb}%
  \BibitemOpen
  \bibfield  {author} {\bibinfo {author} {\bibfnamefont {A.}~\bibnamefont
  {D{\"{o}}rnen}}, \bibinfo {author} {\bibfnamefont {G.}~\bibnamefont {Pensl}},
  \ and\ \bibinfo {author} {\bibfnamefont {R.}~\bibnamefont {Sauer}},\ }\href
  {\doibase 10.1103/PhysRevB.33.1495} {\bibfield  {journal} {\bibinfo
  {journal} {Phys. Rev. B}\ }\textbf {\bibinfo {volume} {33}},\ \bibinfo
  {pages} {1495(R)} (\bibinfo {year} {1986}{\natexlab{a}})}\BibitemShut
  {NoStop}%
\bibitem [{\citenamefont {D{\"{o}}rnen}\ \emph
  {et~al.}(1986{\natexlab{b}})\citenamefont {D{\"{o}}rnen}, \citenamefont
  {Pensl},\ and\ \citenamefont {Sauer}}]{ref:dornen86}%
  \BibitemOpen
  \bibfield  {author} {\bibinfo {author} {\bibfnamefont {A.}~\bibnamefont
  {D{\"{o}}rnen}}, \bibinfo {author} {\bibfnamefont {G.}~\bibnamefont {Pensl}},
  \ and\ \bibinfo {author} {\bibfnamefont {R.}~\bibnamefont {Sauer}},\ }\href
  {\doibase 10.1016/0038-1098(86)90167-5} {\bibfield  {journal} {\bibinfo
  {journal} {Solid State Commun.}\ }\textbf {\bibinfo {volume} {57}},\ \bibinfo
  {pages} {861} (\bibinfo {year} {1986}{\natexlab{b}})}\BibitemShut {NoStop}%
\bibitem [{\citenamefont {D{\"{o}}rnen}\ \emph {et~al.}(1987)\citenamefont
  {D{\"{o}}rnen}, \citenamefont {Pensl},\ and\ \citenamefont
  {Sauer}}]{ref:dornen87}%
  \BibitemOpen
  \bibfield  {author} {\bibinfo {author} {\bibfnamefont {A.}~\bibnamefont
  {D{\"{o}}rnen}}, \bibinfo {author} {\bibfnamefont {G.}~\bibnamefont {Pensl}},
  \ and\ \bibinfo {author} {\bibfnamefont {R.}~\bibnamefont {Sauer}},\ }\href
  {\doibase 10.1103/PhysRevB.35.9318} {\bibfield  {journal} {\bibinfo
  {journal} {Phys. Rev. B}\ }\textbf {\bibinfo {volume} {35}},\ \bibinfo
  {pages} {9318} (\bibinfo {year} {1987})}\BibitemShut {NoStop}%
\bibitem [{\citenamefont {D{\"{o}}rnen}\ \emph {et~al.}(1988)\citenamefont
  {D{\"{o}}rnen}, \citenamefont {Sauer},\ and\ \citenamefont
  {Pensl}}]{ref:dornen88}%
  \BibitemOpen
  \bibfield  {author} {\bibinfo {author} {\bibfnamefont {A.}~\bibnamefont
  {D{\"{o}}rnen}}, \bibinfo {author} {\bibfnamefont {R.}~\bibnamefont {Sauer}},
  \ and\ \bibinfo {author} {\bibfnamefont {G.}~\bibnamefont {Pensl}},\ }\href
  {\doibase 10.1007/BF02652141} {\bibfield  {journal} {\bibinfo  {journal} {J.
  Electron. Mater.}\ }\textbf {\bibinfo {volume} {17}},\ \bibinfo {pages} {121}
  (\bibinfo {year} {1988})}\BibitemShut {NoStop}%
\bibitem [{\citenamefont {Davies}(1989)}]{ref:davies89}%
  \BibitemOpen
  \bibfield  {author} {\bibinfo {author} {\bibfnamefont {G.}~\bibnamefont
  {Davies}},\ }\href {\doibase 10.1016/0370-1573(89)90064-1} {\bibfield
  {journal} {\bibinfo  {journal} {Phys. Rep.}\ }\textbf {\bibinfo {volume}
  {176}},\ \bibinfo {pages} {83} (\bibinfo {year} {1989})}\BibitemShut
  {NoStop}%
\bibitem [{\citenamefont {Armelles}\ \emph {et~al.}(1986)\citenamefont
  {Armelles}, \citenamefont {Barrau}, \citenamefont {Thomas},\ and\
  \citenamefont {Brousseau}}]{ref:armelles1986}%
  \BibitemOpen
  \bibfield  {author} {\bibinfo {author} {\bibfnamefont {G.}~\bibnamefont
  {Armelles}}, \bibinfo {author} {\bibfnamefont {J.}~\bibnamefont {Barrau}},
  \bibinfo {author} {\bibfnamefont {V.}~\bibnamefont {Thomas}}, \ and\ \bibinfo
  {author} {\bibfnamefont {M.}~\bibnamefont {Brousseau}},\ }\href {\doibase
  10.1088/0022-3719/19/14/026} {\bibfield  {journal} {\bibinfo  {journal} {J.
  Phys. C Solid State Phys.}\ }\textbf {\bibinfo {volume} {19}},\ \bibinfo
  {pages} {2593} (\bibinfo {year} {1986})}\BibitemShut {NoStop}%
\bibitem [{\citenamefont {Sommerfeld}(1931)}]{ref:sommerfeld}%
  \BibitemOpen
  \bibfield  {author} {\bibinfo {author} {\bibfnamefont {A.}~\bibnamefont
  {Sommerfeld}},\ }\href {\doibase https://doi.org/10.1002/andp.19314030302}
  {\bibfield  {journal} {\bibinfo  {journal} {Ann. Phys.}\ }\textbf {\bibinfo
  {volume} {403}},\ \bibinfo {pages} {257} (\bibinfo {year}
  {1931})}\BibitemShut {NoStop}%
\bibitem [{\citenamefont {Turiansky}\ \emph {et~al.}(2024)\citenamefont
  {Turiansky}, \citenamefont {Alkauskas},\ and\ \citenamefont {{Van de
  Walle}}}]{ref:turiansky_jpcm2024}%
  \BibitemOpen
  \bibfield  {author} {\bibinfo {author} {\bibfnamefont {M.~E.}\ \bibnamefont
  {Turiansky}}, \bibinfo {author} {\bibfnamefont {A.}~\bibnamefont
  {Alkauskas}}, \ and\ \bibinfo {author} {\bibfnamefont {C.~G.}\ \bibnamefont
  {{Van de Walle}}},\ }\href {\doibase 10.1088/1361-648X/ad2588} {\bibfield
  {journal} {\bibinfo  {journal} {J. Phys. Condens. Matter}\ }\textbf {\bibinfo
  {volume} {36}},\ \bibinfo {pages} {195902} (\bibinfo {year}
  {2024})}\BibitemShut {NoStop}%
\bibitem [{\citenamefont
  {Griffiths}(2005{\natexlab{b}})}]{ref:griffiths_hyd-1s-wf}%
  \BibitemOpen
  \bibfield  {author} {\bibinfo {author} {\bibfnamefont {D.~J.}\ \bibnamefont
  {Griffiths}},\ }\href@noop {} {\emph {\bibinfo {title} {{Introduction to
  Quantum Mechanics}}}},\ \bibinfo {edition} {2nd}\ ed.\ (\bibinfo  {publisher}
  {Pearson Education},\ \bibinfo {year} {2005})\ p.\ \bibinfo {pages}
  {151}\BibitemShut {NoStop}%
\bibitem [{\citenamefont {Dreyer}\ \emph {et~al.}(2020)\citenamefont {Dreyer},
  \citenamefont {Alkauskas}, \citenamefont {Lyons},\ and\ \citenamefont {{Van
  de Walle}}}]{ref:dreyer}%
  \BibitemOpen
  \bibfield  {author} {\bibinfo {author} {\bibfnamefont {C.~E.}\ \bibnamefont
  {Dreyer}}, \bibinfo {author} {\bibfnamefont {A.}~\bibnamefont {Alkauskas}},
  \bibinfo {author} {\bibfnamefont {J.~L.}\ \bibnamefont {Lyons}}, \ and\
  \bibinfo {author} {\bibfnamefont {C.~G.}\ \bibnamefont {{Van de Walle}}},\
  }\href {\doibase 10.1103/PhysRevB.102.085305} {\bibfield  {journal} {\bibinfo
   {journal} {Phys. Rev. B}\ }\textbf {\bibinfo {volume} {102}},\ \bibinfo
  {pages} {085305} (\bibinfo {year} {2020})}\BibitemShut {NoStop}%
\bibitem [{\citenamefont {Kazemi}\ \emph {et~al.}(2025)\citenamefont {Kazemi},
  \citenamefont {Keshavarz}, \citenamefont {Turiansky}, \citenamefont {Lyons},
  \citenamefont {Abrosimov}, \citenamefont {Simmons}, \citenamefont
  {Higginbottom},\ and\ \citenamefont {Thewalt}}]{ref:kazemi2025}%
  \BibitemOpen
  \bibfield  {author} {\bibinfo {author} {\bibfnamefont {M.}~\bibnamefont
  {Kazemi}}, \bibinfo {author} {\bibfnamefont {M.}~\bibnamefont {Keshavarz}},
  \bibinfo {author} {\bibfnamefont {M.~E.}\ \bibnamefont {Turiansky}}, \bibinfo
  {author} {\bibfnamefont {J.~L.}\ \bibnamefont {Lyons}}, \bibinfo {author}
  {\bibfnamefont {N.~V.}\ \bibnamefont {Abrosimov}}, \bibinfo {author}
  {\bibfnamefont {S.}~\bibnamefont {Simmons}}, \bibinfo {author} {\bibfnamefont
  {D.~B.}\ \bibnamefont {Higginbottom}}, \ and\ \bibinfo {author}
  {\bibfnamefont {M.~L.~W.}\ \bibnamefont {Thewalt}},\ }\href
  {https://arxiv.org/abs/2510.23862} {\enquote {\bibinfo {title} {Giant isotope
  effect on the excited-state lifetime and emission efficiency of the silicon t
  centre},}\ } (\bibinfo {year} {2025}),\ \Eprint
  {http://arxiv.org/abs/2510.23862} {arXiv:2510.23862 [quant-ph]} \BibitemShut
  {NoStop}%
\end{thebibliography}%

\end{document}


\floatsetup[figure]{style=plain,subcapbesideposition=top} 
\floatsetup[table]{style=plain, capposition=top} 

\preprint{APS/123-QED}

\title{Supplemental Material: A CN complex as an alternative to the T center in Si}


\author{J. K. Nangoi}
\email[Corresponding author: nangoi@ucsb.edu]{}
\affiliation{
    Materials Department, University of California, Santa Barbara, California 93106, USA
}

\author{M. E. Turiansky}
\affiliation{
    Materials Department, University of California, Santa Barbara, California 93106, USA
}
\affiliation{
    US Naval Research Laboratory, Washington, DC 20375, USA
}

\author{C. G. Van de Walle}
\affiliation{
    Materials Department, University of California, Santa Barbara, California 93106, USA
}


%



\maketitle


\section{Kohn-Sham states of non-neutral charged states} \label{sec:supp_KS-pm1}

\begin{figure}[h!]
    \centering

    \includegraphics[width=\linewidth]{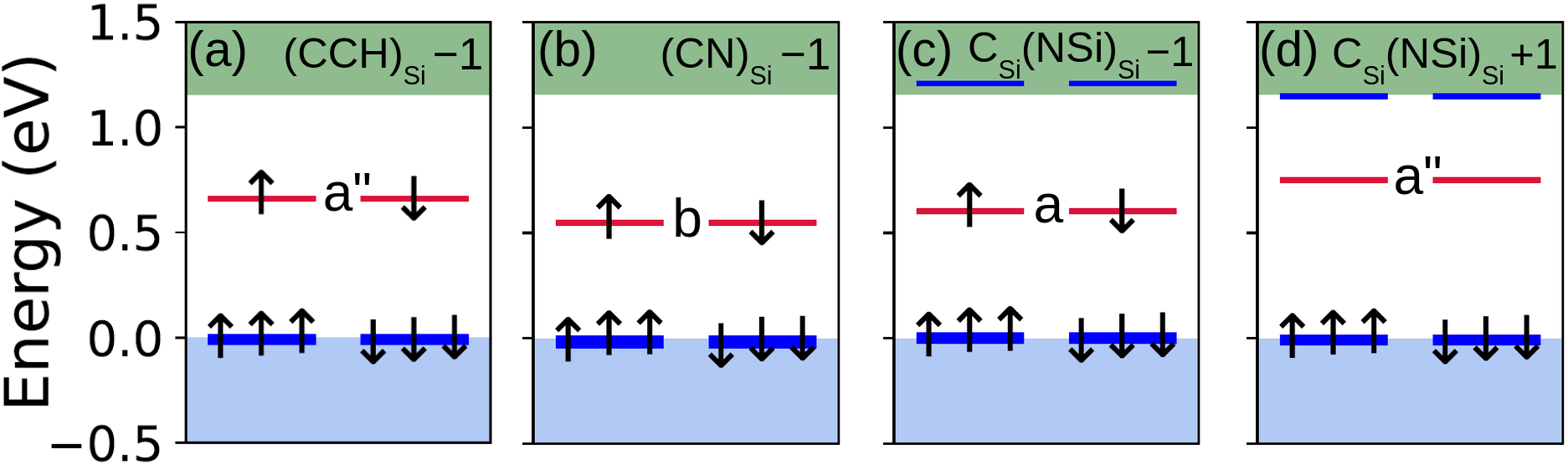}
    
    \caption{
        Ground-state Kohn-Sham states for the non-neutral charged states of the T center [(a)] and the CN defects [(b)--(d)].
    }

    \label{fig:KS-eigenvals_pm1}
\end{figure}

\FloatBarrier 

\vspace{-2em}
\section{Other defects considered} 
\label{sec:supp_other-defects}

\begin{figure}[h!]
    \centering
    
    \includegraphics[width=\linewidth]{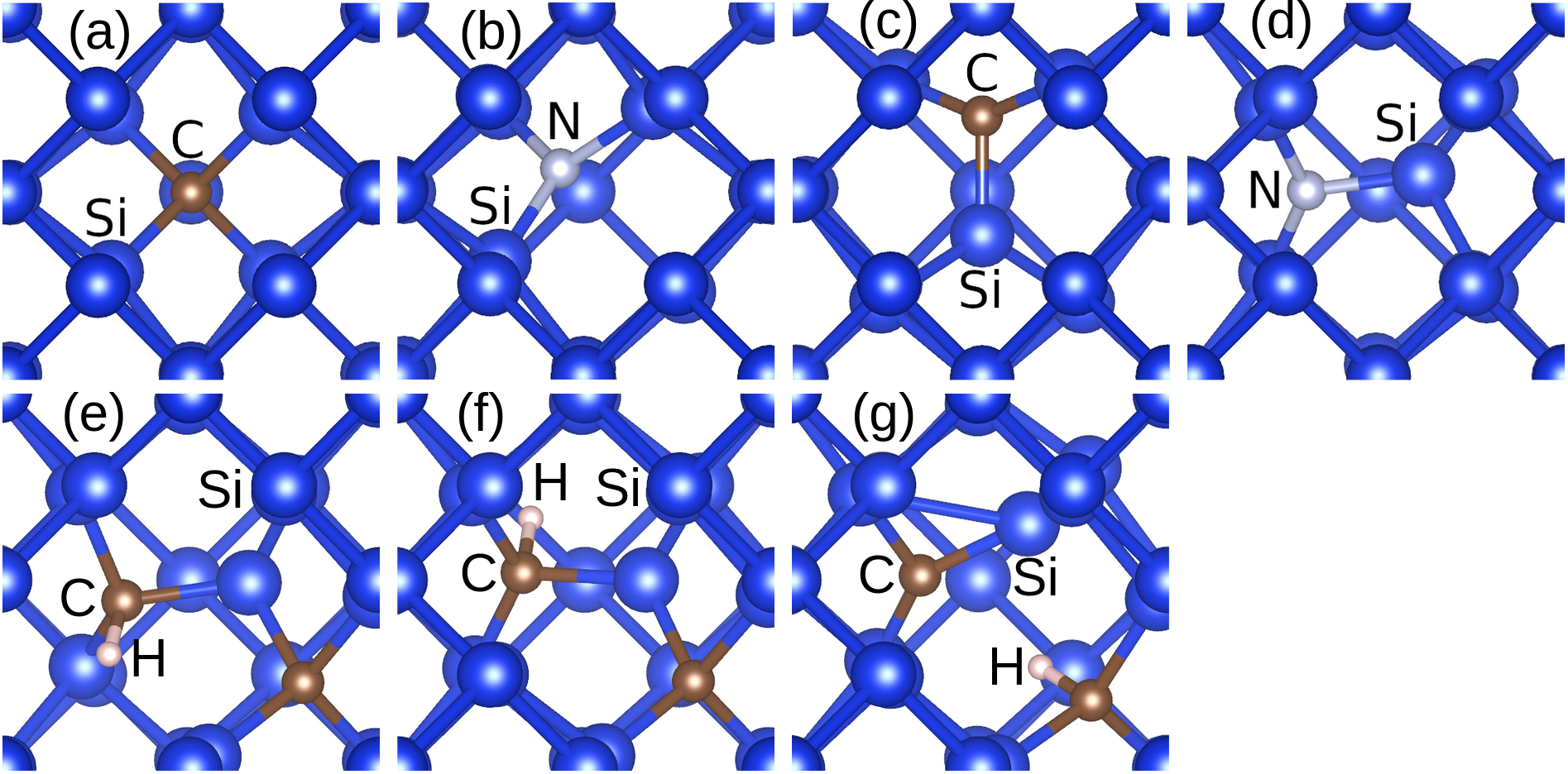}

    \caption{
        Atomic structure, in the neutral charge state, of (a) C$_\mathrm{Si}$, (b) N$_\mathrm{Si}$, (c) (CSi)$_\mathrm{Si}$, 
        (d) (NSi)$_\mathrm{Si}$, 
        (e) C$_\mathrm{Si}$(CHSi)$_\mathrm{Si}$ 1, 
        (f) C$_\mathrm{Si}$(CHSi)$_\mathrm{Si}$ 2, and 
        (g) (CSi)$_\mathrm{Si}$(CH)$_\mathrm{Si}$. 
    }

    \label{fig:struct_others}
\end{figure}

\begin{figure}[h!]
    \centering
    \includegraphics[width=\linewidth]{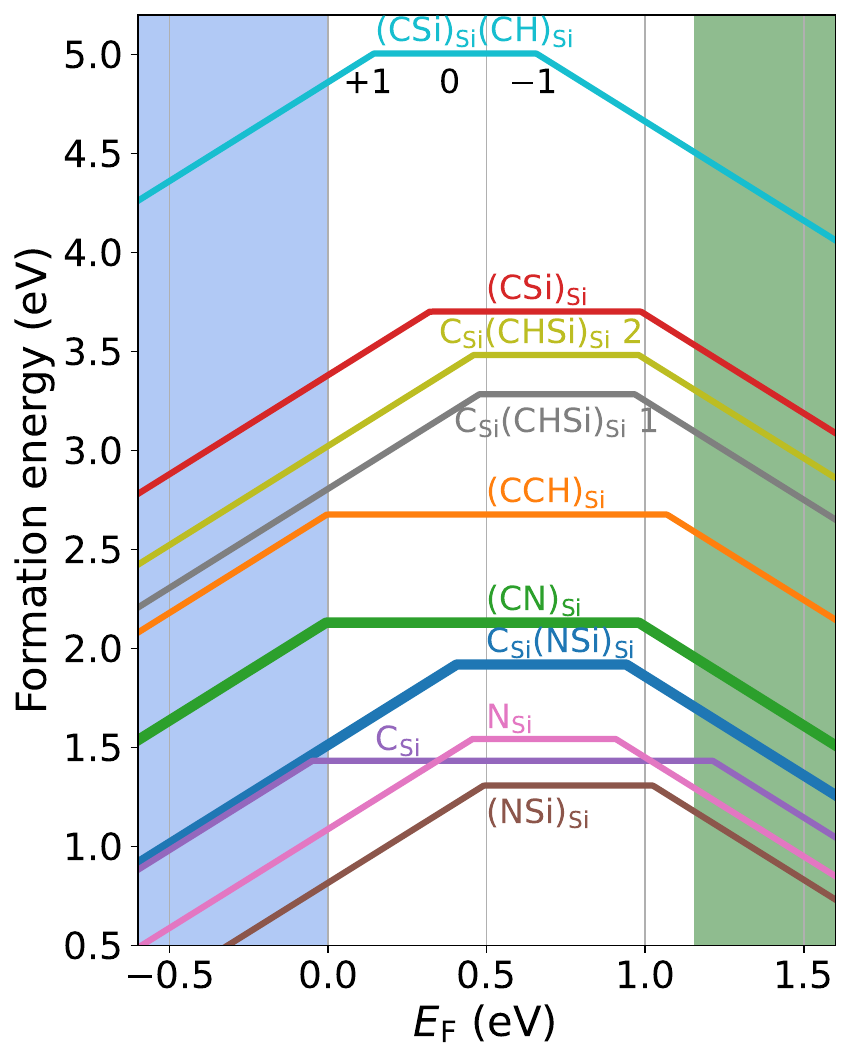}

    \caption{
        Defect formation energies as functions of Fermi level for the $+1$, 0, and $-1$ charge states of all defects considered. HSE functional and supercell size of $4 \times 4 \times 4$ (512 atoms) are used.
    }

    \label{fig:Eform_all}
\end{figure}

Figures {\ref{fig:struct_others}}(a)--(d) show the lowest-energy  structures of C$_\mathrm{Si}$, N$_\mathrm{Si}$, C$_i$, and N$_i$ in Si used in this work, consistent with previous first-principles or molecular dynamics calculations~{\cite{ref:tersoff1990, ref:platonenko_N-int, ref:platonenko_C-sub, ref:platonenko_C-int, ref:simha2023}}.  
Their formation energy diagrams are shown in Fig.~{\ref{fig:Eform_all}}.

Because C$_\mathrm{Si}$(NSi)$_\mathrm{Si}$ [Fig.~{\ref{fig:struct}}(c) in the main text] has lower formation energy than (CN)$_\mathrm{Si}$, we also explore other analogous structures of CCH complex to see if they are more stable than the T center structure (CCH)$_\mathrm{Si}$. We replace N in C$_\mathrm{Si}$(NSi)$_\mathrm{Si}$ with C, and place H at various positions: 
(1) near (CSi)$_\mathrm{Si}$, 
(2) near C$_\mathrm{Si}$, and 
(3) between the 2 carbons. 
After relaxation, 
\#(1) becomes two distinct structures of C$_\mathrm{Si}$(CHSi)$_\mathrm{Si}$, labeled 1 and 2 [Figs. {\ref{fig:struct_others}}(e) and (f)], 
\#(2) becomes (CSi)$_\mathrm{Si}$(CH)$_\mathrm{Si}$ [Fig.~{\ref{fig:struct_others}}(g)], and 
\#(3) becomes C$_\mathrm{Si}$(CHSi)$_\mathrm{Si}$ 1 [Fig.~{\ref{fig:struct_others}}(e)]. 
As seen in the formation energy diagram [Fig.~{\ref{fig:Eform_all}}], in the neutral charge state, these structures are actually 0.6--2.5~eV higher in energy than (CCH)$_\mathrm{Si}$.

\FloatBarrier 

\section{Migration barrier of \texorpdfstring{(NS\MakeLowercase{i})$_\mathrm{Si}$}{}} \label{sec:supp_neb}

As shown in the main text, the lowest-energy decomposition reaction of (CN)$_\mathrm{Si}^0$ and [C$_\mathrm{Si}$(NSi)$_\mathrm{Si}$]$^0$ produces C$_\mathrm{Si}^0$ + (NSi)$_\mathrm{Si}^0$, with (endothermic) decomposition energies of 0.61~eV and 0.82~eV. 
To estimate the barriers for these decomposition reactions, we compute the migration barrier of the interstitial product, namely (NSi)$_\mathrm{Si}^0$.

\begin{figure}[h!]
    \centering
    \includegraphics[width=\linewidth]{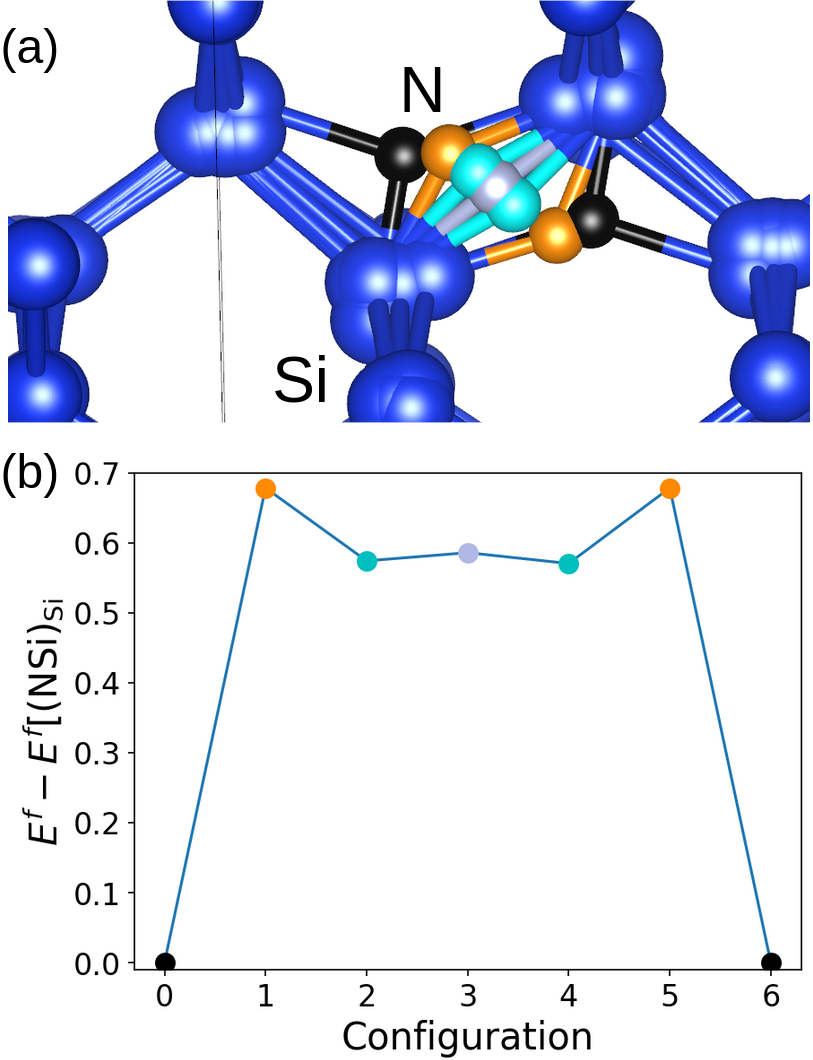}

    \caption{
        (a) Migration pathway of (NSi)$_\mathrm{Si}^0$. 
        Blue circles are Si. 
        Black circles are nitrogens in equivalent (NSi)$_\mathrm{Si}^0$ configurations; 
        cyan, bond-centered N$_i^0$;  
        gray, intermediate;
        orange, highest-energy. 
        (b) Formation energy differences with respect to that of (NSi)$_\mathrm{Si}^0$. 
        Circles label the configurations shown in panel (a), drawn with the same colors. 
    }

    \label{fig:neb}
\end{figure}

Reference~{\onlinecite{ref:schultz2001}} suggested a pathway for migration of (NSi)$_\mathrm{Si}^0$: the N moves between two equivalent (NSi)$_\mathrm{Si}^0$ sites through the bond-centered site. We therefore perform climbing image nudged elastic band (CI-NEB) 
calculations~{\cite{ref:Henkelman2000_climbing-neb}} between our calculated lowest-energy configuration of (NSi)$_\mathrm{Si}^0$ [Fig.~{\ref{fig:neb}}(a), N as black atoms] and the bond-centered configuration of N$_i^0$ [Fig.~{\ref{fig:neb}}(a), N as cyan atoms], finding a saddle-point configuration indicated by N as orange atoms. 
We also perform CI-NEB between 2 neighboring bond-centered configurations (which are equivalent by symmetry), finding a saddle-point configuration for which the N is in gray.
In the figure, the atomic sites with the same color are equivalent by symmetry. 
Figure~{\ref{fig:neb}}(b) shows the corresponding reaction pathway, for which the configurations are color-coded the same way as in Fig.~{\ref{fig:neb}}(a). The figure shows that the barrier is 0.68~eV.

The results above use HSE with a supercell size of $3 \times 3 \times 3$ (216 atoms) which is sufficiently large to converge the reaction barrier within 0.03~eV as tested using PBE. 
The saddle points are calculated using 1-image CI-NEB (i.e., 1 intermediate configuration between the starting and ending configurations). 
Using PBE, we have tested that the 1-image calculation results in a saddle-point configuration whose energy is within $2 \times 10^{-5}$~eV from that obtained using 3-image CI-NEB. 

We compute configurations 0 and 2 in Fig.~{\ref{fig:neb}}(b) using density functional theory (DFT), and then perform 1-image CI-NEB between those two, finding configuration 1. Then, we rotate configuration 2 to construct configuration 4, and perform 1-image CI-NEB between those two, finding configuration 3. 
Finally, path 4 $\to$ 6 is equivalent by symmetry to 0 $\leftarrow$ 2.

\FloatBarrier 

\section{Bound exciton wavefunction and supercell-size dependence} \label{sec:supp_sup-dep}

As discussed in the main text, the excited state consists of a defect-bound exciton consisting of a hydrogenic hole (electron) bound to an electron (hole) localized at the defect site. 
As an approximation, we can describe the hydrogenic charge's wavefunction within effective mass theory~{\cite{ref:wannier}}. 
Following the ansatz of Kohn and Luttinger~{\cite{ref:kohn-luttinger}}, the hydrogenic 
wavefunction takes the form~{\cite{ref:passler}}
\begin{align}
    \psi_h(\mathbf{r}) = \sqrt{\mathcal{N}_0 \Omega_0} ~\phi(\mathbf{r}) ~u_{\mathbf{k}_0} (\mathbf{r}), \label{eqn:supp_deloc-wf}
\end{align}
where 
$\mathcal{N}_0$ is the number of unit cells (each with volume $\Omega_0$) that the wavefunction extends over, 
$u_{\mathbf{k}_0}$ is the unperturbed lattice-periodic part of the Bloch function of the crystal (at the k-point $\mathbf{k}_0$ where the band extremum is located, i.e., VBM for the hydrogenic hole and CBM for the hydrogenic electron) normalized over a single unit cell, and 
$\phi$ is the envelope function that satisfies the Wannier equation~{\cite{ref:wannier, ref:passler}}, which is normalized over the entire volume $\mathcal{N}_0 \Omega_0$. Taking $\mathcal{N}_0$ to infinity, the 
negative-energy (bound-state) solutions to the Wannier equation are the hydrogen wavefunctions, scaled appropriately by the band effective mass $m^*$ and bulk dielectric constant $\epsilon_r$. The corresponding effective Bohr radius $a_0^*$ that characterizes $\phi$ is then given by 
\begin{align}
    a_0^* = \frac{4\pi\epsilon_0\hbar^2}{e^2} \frac{\epsilon_r}{m^*}. \label{eqn:supp_bohr-rad}
\end{align}

\begin{figure*}[]
    \centering

    \includegraphics[width=\linewidth]{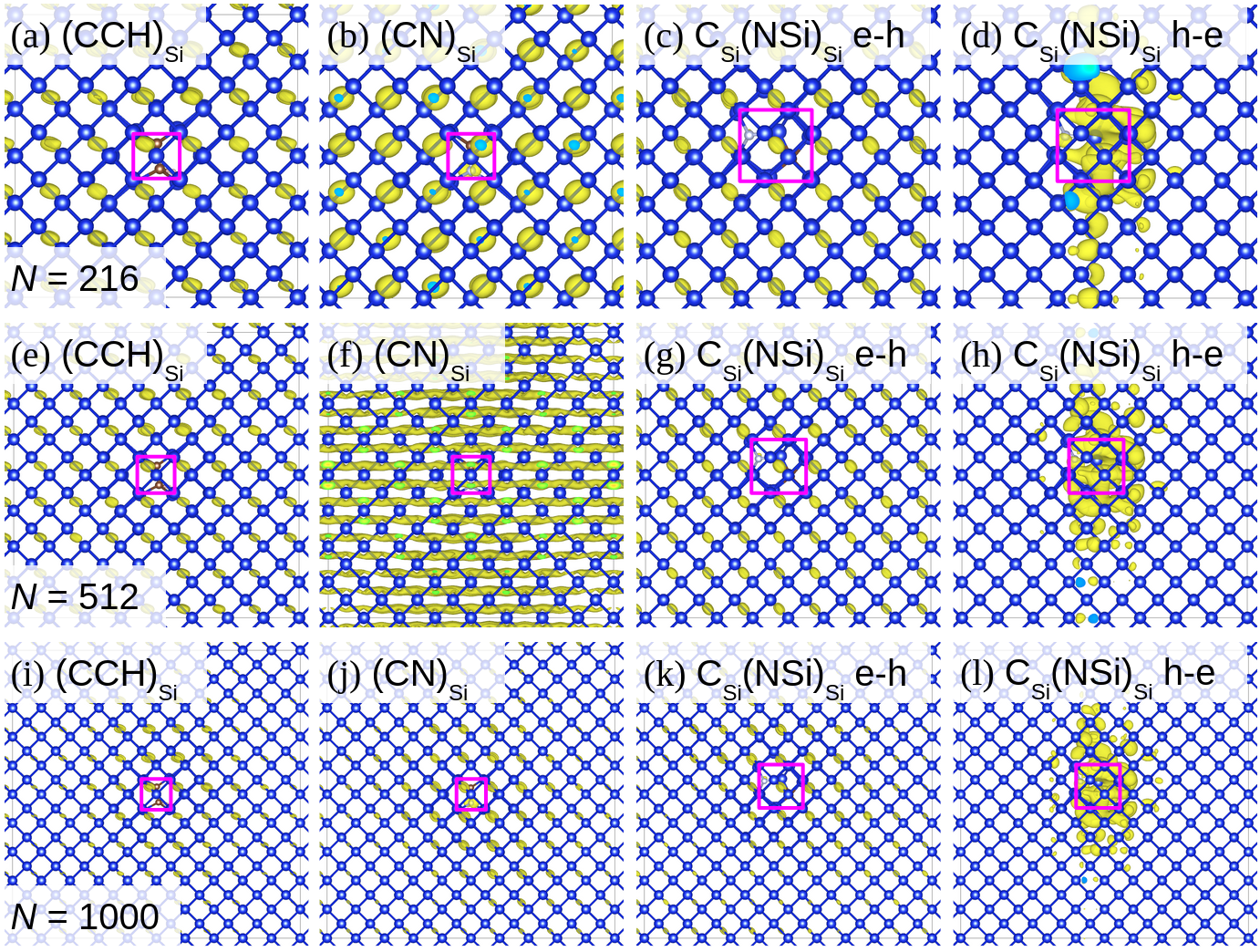}
    
    \caption{
        Isosurfaces (yellow) of real-space Kohn-Sham probability densities, 
        shown for the whole supercell, corresponding to the \emph{hydrogenic} charge in the electronic excited state 
        [(d), (h), (l) for hydrogenic electron ``h-e''; rest for hydrogenic hole ``e-h'']
        for supercell size $N$ = 216 (top row), 512 (middle row), and 1000 (bottom row). 
        Here $N$ denotes the number of Si atoms in the defect-free supercell. 
        Blue circles are Si; brown, C; light blue, N; and pink, H. 
        Purple squares indicate the defect sites. 
        The isosurface levels used are 
        55.16\%, 
        14.23\%, 
        60.15\% and 6.42\% 
        of the maximum charge density in each supercell for 
        (CCH)$_\mathrm{Si}$, 
        (CN)$_\mathrm{Si}$, 
        C$_\mathrm{Si}$(NSi)$_\mathrm{Si}$ e-h and h-e, 
        except for panel (f), for which we use 51.15\% (because 14.23\% is too small and yields an isosurface that fills the entire supercell).
    }

    \label{fig:chgDens_bound}
\end{figure*}

For the case of the exciton with a hydrogenic hole, 
we expect that, as an approximation, the heavy (as opposed to the light) hole makes up the bound exciton, because the Rydberg energy~{\cite{ref:griffiths_rydberg}}, which approximates the excitonic binding energy, is proportional to the effective mass $m^*$, and therefore the heavy hole will have larger binding energy. 
Using the experimental values for the Si heavy hole effective mass of 0.49~$m_0$~{\cite{ref:si_exp-lattConst}} 
(where $m_0$ is the free-electron mass) 
and dielectric constant of 11.7~{\cite{ref:Si_dielConst}} yields 
an effective Bohr radius $a_0^*$ [Eq.~{\eqref{eqn:supp_bohr-rad}}] of $\sim$13~{\AA} for the heavy hole. 
(For the T center, whose experimental binding energy is 35~meV~{\cite{ref:bergeron}}, the corresponding $m^* = 0.35m_0$ is indeed closer to the heavy hole mass than to the light hole mass of 0.16~$m_0$~{\cite{ref:si_exp-lattConst}}).
The length of the largest supercell size we have considered, $5 \times 5 \times 5$ conventional cubic cells (1000 Si atoms), is $\sim$27~\AA, 
around twice $a_0^*$. 
As seen in Figs.~{\ref{fig:chgDens_bound}}(a)--(c), (e)--(g), and (i)--(k), the hole seems to barely fit in the largest supercell, and appears delocalized over practically the entire supercell for the two smaller ones.

For the case of the exciton with a hydrogenic electron, 
we use an effective mass of 0.98~$m_0$, equal to the longitudinal mass of the Si CBM (which is larger than the transverse mass 0.19~$m_0$)~{\cite{ref:si_exp-lattConst}}, 
corresponding to an effective Bohr radius of $\sim$6~{\AA}. 
We choose the longitudinal mass because, as discussed above, larger effective mass means higher excitonic binding energy. 
Similar to the hydrogenic-hole case, here we find that the electron seems to barely fit in the largest supercell, and appears to extend beyond the supercell for the two smaller ones.

\FloatBarrier 

\section{Huang-Rhys factor calculations} \label{sec:supp_HR}

\begin{figure}[h!]
    \centering

    \includegraphics[width=\linewidth]{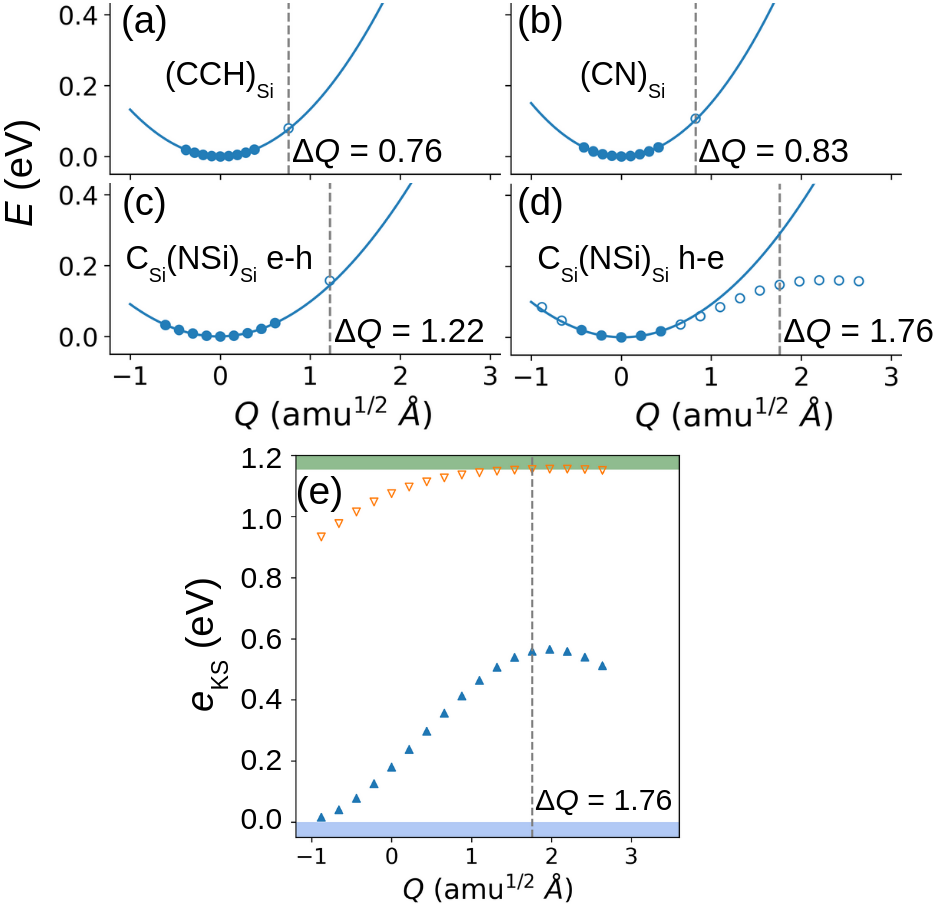}
    
    \caption{
        (a)--(d) Configuration coordinate curves of the electronic ground state as a function of one-dimensional coordinate $Q$. Circles are DFT-calculated total energies $E$ referenced to the energy of the equilibrium structures ($Q=0$); only filled circles are used in the parabolic fits (blue curves). 
        (e) Kohn-Sham eigenvalues of the $a$ defect states for C$_\mathrm{Si}$(NSi)$_\mathrm{Si}$ as a function of Q: blue = occupied, orange = unoccupied. 
        The vertical dashed line indicates the $\Delta Q$ value for the equilibrium structure of the electronic excited state. 
    }

    \label{fig:HR_CCD}
\end{figure}

The Huang-Rhys factor is defined as $S = E_\mathrm{r} / (\hbar \Omega)$ where $E_\mathrm{r}$ equals the difference between the 
ground-state 
energy at the equilibrium structure of the \emph{excited} state and that of the \emph{ground} state, 
and $\Omega$ is the phonon frequency in the ground state within the one-dimensional approximation~{\cite{ref:alkauskas2012_huang-rhys}}; the values of both are reported in Table~{\ref{tab:huang-rhys}} in the main text. 
Figure~{\ref{fig:HR_CCD}} shows the one-dimensional configuration-coordinate curves for the T center and the CN defects in the electronic ground state. 
$\Omega$ is calculated from the parabolic fit $E = (1/2)\Omega^2Q^2$~{\cite{ref:alkauskas2012_huang-rhys}} to the energies calculated using DFT. 

For the T center, (CN)$_\mathrm{Si}$, and C$_\mathrm{Si}$(NSi)$_\mathrm{Si}$ localized-electron case ``e-h'' [Figs.~{\ref{fig:HR_CCD}}(a)--(c)], the DFT-calculated potential energy surface is harmonic until at least the equilibrium excited-state structure. Therefore, $E_\mathrm{r}$ equals $E$ indicated by the open circle.

For the C$_\mathrm{Si}$(NSi)$_\mathrm{Si}$ localized-hole case ``h-e'' [Fig.~{\ref{fig:HR_CCD}}(d)], the DFT-calculated potential energy surface appears anharmonic. 
Figure~{\ref{fig:HR_CCD}}(e) shows the Kohn-Sham eigenvalues of the $a$ states shown in Fig.~{\ref{fig:KS-eigenvals}}(c) as functions of $Q$. 
We see that the unoccupied $a$ state crosses the CBM at $Q \approx 1$, affecting the behavior of the occupied $a$ state in the gap in such a way that it becomes nonmonotonic. 
This is consistent with the observation that the apparent potential energy surface in Fig.~{\ref{fig:HR_CCD}}(d) is anharmonic for $Q \geq 1$. 
Data points in the anharmonic regime (open circles) are therefore ignored, and we take $E_\mathrm{r} = (1/2)\Omega^2 (\Delta Q)^2$, where $\Delta Q$ is the $Q$ for the equilibrium structure of the excited state (Fig.~{\ref{fig:HR_CCD}}).

\FloatBarrier 

\vspace{-1.1em}
\section{Full vs. single-shot PBE0} \label{sec:supp_PBE0-test}
\vspace{-1.1em}
\begin{table}[h!]
\caption{
    Full vs. single-shot PBE0 results for bulk Si in primitive cell and T center in 216-atom supercell.
    CTL = charge transition level. $\mu^0$ = transition dipole moment in ground state, Eq.~{\eqref{eqn:supp_mu0-sq}}.
    \label{tab:supp_PBE0-full-vs-single-shot}
}
\begin{ruledtabular}
\begin{tabular}{lcc}
    & Full & Single-shot \\ 
    \hline
    Band gap (eV; bulk Si)               &   1.23  &  1.23 \\
    Dielectric constant (bulk Si)        &  11.736 &  11.738 \\
    \hline
    $E_\mathrm{ZPL}$ (meV)                      & 680     & 681 \\
    $E^\mathrm{f}$, neutral charge state (eV)   &   2.48  &  2.47 \\
    $E_\mathrm{CTL, 0/-} - E_\mathrm{VBM}$ (eV) &   0.93  &  0.93 \\
    $\mu^0$ (e{\AA})           &   0.201 &  0.195 \\
\end{tabular}
\end{ruledtabular}
\end{table}

\FloatBarrier 

\section{Zero-phonon line of all centers} \label{sec:supp_ZPL}

\begin{figure}[h!]
    \centering
    \includegraphics[width=\linewidth]{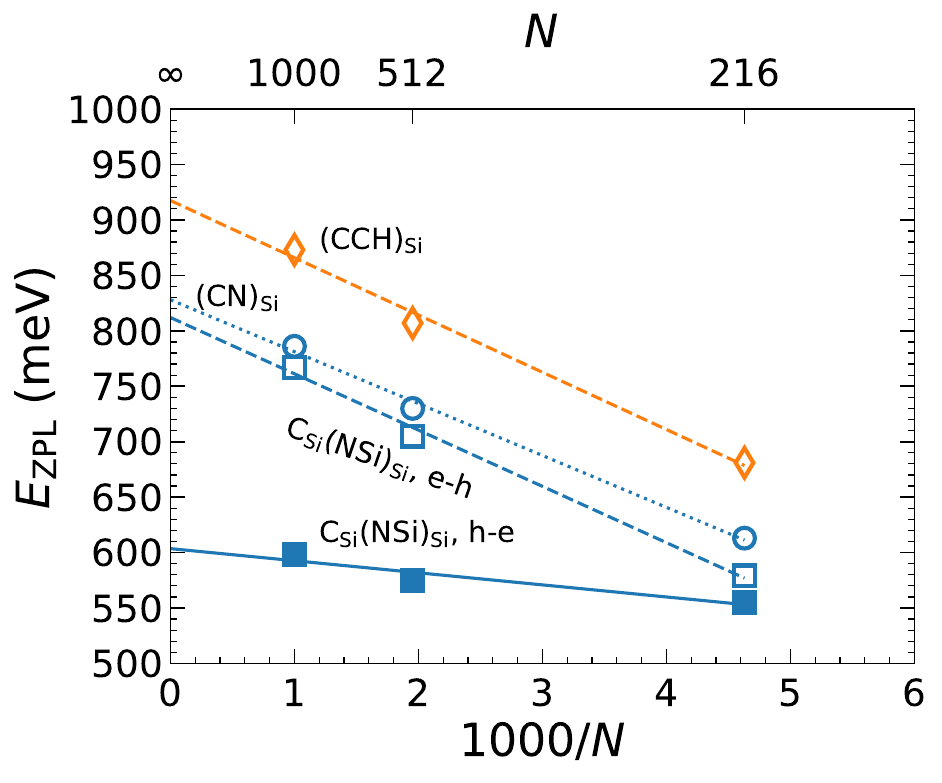}
    \caption{
        ZPL energies for the T and all CN centers calculated using single-shot PBE0.
    }

    \label{fig:supp_supdep_ZPL}
\end{figure}

Figure {\ref{fig:supp_supdep_ZPL}} shows that the extrapolated PBE0 ZPLs are 812~meV and 604~meV for C$_\mathrm{Si}$(NSi)$_\mathrm{Si}$ localized-electron and localized-hole case. 
Both of these are far from the observed ZPL range of 746 to 772~meV attributed to complexes containing C and N (and possibly O)~\cite{ref:dornen85,ref:dornen86-prb,ref:dornen86,ref:dornen87,ref:dornen88,ref:davies89}. 
A few luminescence lines close to 812~meV have been observed in Si, at 811~meV~{\cite{ref:armelles1986}} and 829.8~meV~{\cite{ref:davies89}}.

\FloatBarrier 

\section{Radiative Lifetime for a Bound Exciton Emitter} \label{sec:supp_mu}

\begin{figure}[]
    \centering

    \includegraphics[width=\linewidth]{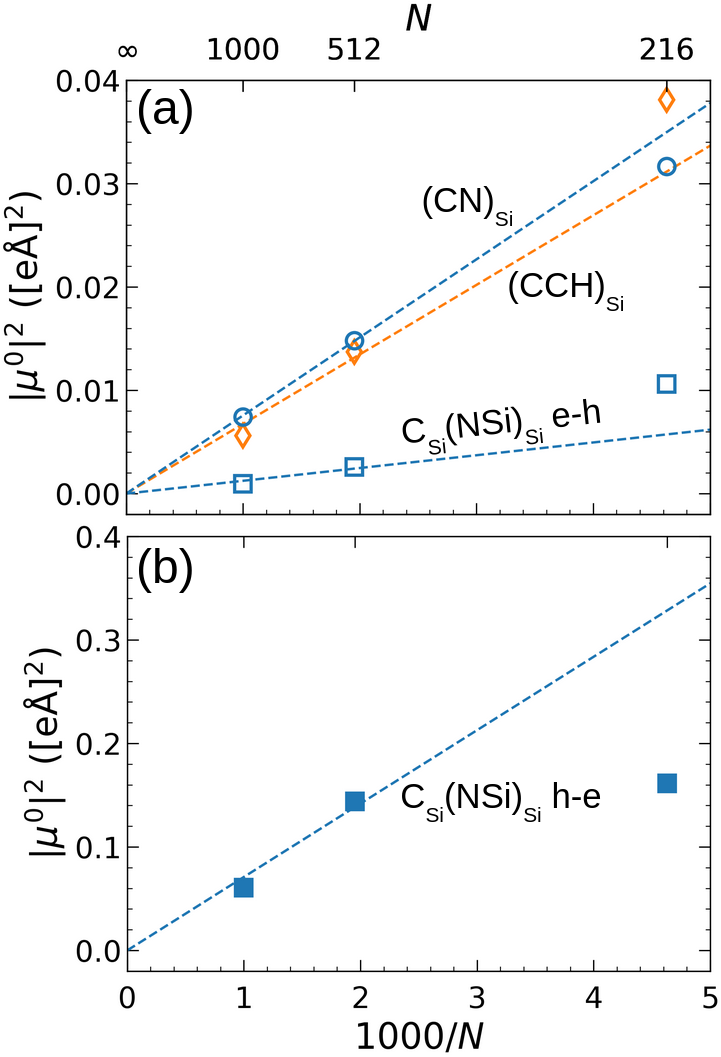}
    \caption{
        Modulus square of the transition dipole moment $|\mathbf{\mu}^0|^2$ [Eq.~{\eqref{eqn:supp_mu0-sq}}] as a function of supercell size. 
        Lines are linear fits constrained to the origin excluding $N =216$.
    }

    \label{fig:supdep_dipol}
\end{figure}

The transition dipole moment $\mu$ that enters the radiative lifetime $\tau$, Eq.~{\eqref{eqn:radiative-lifetime}}, is given by
\begin{align}
    |\mathbf{\mu}|^2 = |\braket{\psi_l | e \mathbf{r} | \psi_h}|^2, \label{eqn:supp_mu-sq}
\end{align}
where $e$ is the elementary charge, $\mathbf{r}$ is the position operator, 
$\psi_l$ is the wavefunction of the charge localized at the defect, and $\psi_h$ is the hydrogenic wavefunction given by Eq.~{\eqref{eqn:supp_deloc-wf}}. 
Evaluating $|\mu|^2$ explicitly is challenging because $\psi_h$ can extend over distances larger than computationally tractable supercell sizes, as discussed in Sec.~{\ref{sec:supp_sup-dep}}. 
A more convenient quantity to evaluate is the transition dipole moment $\mathbf{\mu}^0$ 
in the ground state (for which there is no bound exciton), 
\begin{align}
    |\mathbf{\mu}^0|^2 = |\braket{\psi_l | e \mathbf{r} | \psi_h^0}|^2, \label{eqn:supp_mu0-sq}
\end{align}
where $\psi_h^0$ is 
the free-carrier (Bloch) wavefunction and is given by Eq.~{\eqref{eqn:supp_deloc-wf}} with an envelope function $\phi^0 (\mathbf{r}) = \exp{(i \mathbf{k}_0 \cdot \mathbf{r})} / \sqrt{\mathcal{N}_0 \Omega_0}$. 
Because $\psi_h^0$ is delocalized while $\psi_l$ is localized, we expect $|\mathbf{\mu}^0|^2 \propto 1/(\mathcal{N}_0 \Omega_0)$, and we later confirm this with explicit calculations of $\mu^0$ (Fig.~{\ref{fig:supdep_dipol})}.

To get $\mu$, we introduce the approximation
\begin{align}
    |\mathbf{\mu}|^2 \approx \frac{|\phi(0)|^2}{|\phi^0(0)|^2} |\mathbf{\mu}^0|^2 = \mathcal{N}_0 \Omega_0 |\phi(0)|^2 |\mathbf{\mu}^0|^2 \equiv t |\mathbf{\mu}^0|^2, \label{eqn:supp_mu-sq_approx}
\end{align}
where we have assumed that the localized-charge orbital is centered at the origin. We have introduced the dimensionless scaling parameter $t$ in direct analogy to the Sommerfeld parameter~{\cite{ref:passler, ref:sommerfeld, ref:turiansky_jpcm2024}}. 
For the lowest-energy hydrogenic state, $|\phi(0)|^2 = 1 / [\pi (a_0^*)^3]$~{\cite{ref:griffiths_hyd-1s-wf}}, where $a_0^*$ is the effective Bohr radius [Eq.~{\eqref{eqn:supp_bohr-rad}}]. 
The corresponding $t$ is then
\begin{align}
    t = \mathcal{N}_0 \Omega_0 / [\pi (a_0^*)^3]. \label{eqn:supp_t}
\end{align}
This approximation assumes that 
the orbital character of the hydrogenic hole/electron is 
not significantly changed from the band-edge character; i.e.,
the main effect of binding is to change the density of the hole/electron near the defect. 
Since the product of $t$ and $|\mu^0|^2$ is independent of the volume $\mathcal{N}_0 \Omega_0$, we are free to evaluate it in our supercell volume $\tilde{V}$ (i.e., we set $\mathcal{N}_0 \Omega_0 = \tilde{V}$). 
[Using $\mu$ from Eq.~{\eqref{eqn:supp_mu-sq_approx}} in the expression for radiative lifetime Eq.~{\eqref{eqn:radiative-lifetime}} leads to an expression 
mathematically equivalent to the radiative capture rate 
formalism developed for capture of free carriers by localized defects~{\cite{ref:dreyer}}, as explained in Sec.~{\ref{sec:supp_capture}}.]

Figure~{\ref{fig:supdep_dipol}} shows $\mu^0$ for the T and CN centers 
for different supercell sizes, calculated using single-shot PBE0 as described in the main text. 
These $\mu^0$ values correspond to the transitions between the ground-state Kohn-Sham states discussed in the main text and illustrated in Fig.~{\ref{fig:KS-eigenvals}} there. 
Because both the VBM and CBM contain multiple degenerate bands (heavy, light, and split-off bands for the VB, and 6 valleys for the indirect CBM), 
we project the excited-state band containing the hydrogenic charge to the degenerate ground-state bands, and use the band with the highest projection coefficient to calculate $\mu^0$. 
As seen in Fig.~{\ref{fig:supdep_dipol}}, $|\mu^0|^2$ follows the expected trend of being proportional to 1/N $\propto 1/\tilde{V}$, particularly if we exclude the 216-atom supercell in which the wavefunctions $\psi_l$ are probably not as accurately described. 

Using the slope of $|\mu^0|^2$ with respect to $1/N \propto 1/\tilde{V}$ we obtain $\tilde{V} |\mu^0|^2$.
We then multiply this by $1/[\pi (a_0^*)^3]$ to get $|\mu|^2$ [see Eq.~{\eqref{eqn:supp_mu-sq_approx}}], 
yielding values reported in Table~\ref{tab:supp_muSq-tau}, which includes values for all relevant effective masses (because using the heavy-hole mass, as stated in the main text and Sec.~{\ref{sec:supp_sup-dep}} above, is an approximation).

\begin{table}[h!]
\caption{
    $|\mu|^2$ [Eq.~\eqref{eqn:supp_mu-sq_approx}] and radiative lifetime $\tau$ using various effective masses: holes h, l, $E_b$ = heavy, light, determined from experimental binding energy in Ref.~{\onlinecite{ref:bergeron}}; electrons $\ell$, t = longitudinal, transverse. ``Exp.'' = radiative lifetime deduced from experiments in Ref.~{\onlinecite{ref:kazemi2025}}.
    \label{tab:supp_muSq-tau}
}
\begin{ruledtabular}
\begin{tabular}{lcccccc}
    & \multicolumn{6}{c}{$|\mu|^2$ ([e{\AA}]$^2$)} \\
    & h & l & $E_b$ & Exp. & $\ell$ & t \\
    \hline
    (CCH)$_\mathrm{Si}$                    & 0.0210 & 0.0007 & 0.0077 & - &      - &      - \\
    (CN)$_\mathrm{Si}$                     & 0.0240 & 0.0008 &      - & - &      - &      - \\
    C$_\mathrm{Si}$(NSi)$_\mathrm{Si}$ e-h & 0.0040 & 0.0001 &      - & - &      - &      - \\
    C$_\mathrm{Si}$(NSi)$_\mathrm{Si}$ h-e &      - &      - &      - & - & 1.8080 & 0.0132 \\
    & \multicolumn{6}{c}{$\tau$ ($\mu$s)} \\ 
    \hline
    (CCH)$_\mathrm{Si}$                    &  4.70 & 135 & 12.9 & 4.9 &    - &    - \\
    (CN)$_\mathrm{Si}$                     &  4.18 & 120 &    - &    - &    - &    - \\
    C$_\mathrm{Si}$(NSi)$_\mathrm{Si}$ e-h & 25.55 & 734 &    - &    - &    - &    - \\
    C$_\mathrm{Si}$(NSi)$_\mathrm{Si}$ h-e &     - &   - &    - &    - & 0.06 & 8.23 \\
\end{tabular}
\end{ruledtabular}
\end{table}


\section{Relation to the radiative capture formalism} \label{sec:supp_capture}

Here we note that the radiative lifetime formula with the approximate $\mu$ discussed in the previous section is 
analogous 
to the radiative capture rate formalism~{\cite{ref:dreyer}} developed for capture of a free hole (electron) in valence (conduction) band by a localized defect. 
The difference is that here, 
(1) we have a charge bound to the defect rather than a free carrier, and 
(2) the radiative rate is for each defect-bound exciton, rather than per unit volume of the material containing defects. 

Let $\Gamma_R \equiv 1/\tau$ be the radiative rate given by Eq.~{\eqref{eqn:radiative-lifetime}} in the main text. Using $\mu$ from Eq.~{\eqref{eqn:supp_mu-sq_approx}}, we obtain
\begin{align}
    \Gamma_\mathrm{R} \approx t \Gamma_\mathrm{R}^0 = (t/\tilde{V}) (\tilde{V} \Gamma_\mathrm{R}^0) \equiv \rho 
    C_\mathrm{R}, 
    \label{eqn:supp_radiative-capture-rate}
\end{align}
where $\Gamma_\mathrm{R}^0$ is obtained by substituting $|\mathbf{\mu}^0|^2$ for $|\mathbf{\mu}|^2$ in Eq.~{\eqref{eqn:radiative-lifetime}}. 
As discussed in the previous section, $|\mathbf{\mu}^0|^2$ is inversely proportional to the supercell volume $\tilde{V}$, and therefore $\Gamma_R^0 \propto 1/\tilde{V}$. 
We then define $C_\mathrm{R} \equiv \tilde{V} \Gamma_\mathrm{R}^0$, which is the radiative capture coefficient~{\cite{ref:dreyer}} and is independent of $\tilde{V}$. We also define $\rho \equiv t/\tilde{V}$, which is the effective density of the holes/electrons at the defect, arising from the Coulombic binding of the carrier to the localized charge. For the lowest-energy hydrogenic state, $\rho = [\pi (a_0^*)^3]^{-1}$, equal to the modulus square of the envelope function $|\phi(0)|^2$ we use in Eq.~{\eqref{eqn:supp_mu-sq_approx}}. 

\FloatBarrier 

\bibliography{refs} 